\documentclass[conference]{IEEEtran}
\IEEEoverridecommandlockouts

\usepackage[tight]{subfigure}
\usepackage{graphicx}
\usepackage{hyperref}
\usepackage{booktabs}

\def\BibTeX{{\rm B\kern-.05em{\sc i\kern-.025em b}\kern-.08em
    T\kern-.1667em\lower.7ex\hbox{E}\kern-.125emX}}

\begin{document}

%%
%% The "title" command has an optional parameter,
%% allowing the author to define a "short title" to be used in page headers.
% \title[Preventing TD by TD Aware Project Management]{Preventing Technical Debt by Technical Debt Aware Project Management} 
% \subtitle{Evaluation of a Framework for Managing Technical Debt Developed by Practitioners}
\title{Preventing Technical Debt by Technical Debt Aware Project Management}
%- Evaluation of a Framework for Managing Technical Debt Developed by Practitioners}

%%
%% The "author" command and its associated commands are used to define
%% the authors and their affiliations.

%% Of note is the shared affiliation of the first two authors, and the
%% "authornote" and "authornotemark" commands
%% used to denote shared contribution to the research.

\author{
\IEEEauthorblockN{Marion Wiese}
\IEEEauthorblockA{\textit{Universit{\"a}t Hamburg} \\
%\textit{Universit{\"a}t Hamburg)}\\
Hamburg, Germany \\
wiese@informatik.uni-hamburg.de}
\and
\IEEEauthorblockN{Matthias Riebisch}
\IEEEauthorblockA{\textit{Universit{\"a}t Hamburg} \\
%\textit{name of organization (of Aff.)}\\
Hamburg, Germany \\
riebisch@informatik.uni-hamburg.de}
\and
\IEEEauthorblockN{Julian Schwarze}
\IEEEauthorblockA{\textit{Gruner + Jahr GmbH} \\
Hamburg, Germany \\
schwarze.julian@guj.de}
}

% \author{Marion Wiese}
% \affiliation{%
%   \institution{Universit{\"a}t Hamburg}
%   \city{Hamburg}
%   \country{Germany}
% %\orcid{0000-0003-0160-9531}
% }
% \email{wiese@informatik.uni-hamburg.de}

% \author{Matthias Riebisch}
% \affiliation{%
%   \institution{Universit{\"a}t Hamburg}
%   \city{Hamburg}
%   \country{Germany}
% }
% \email{riebisch@informatik.uni-hamburg.de}

% \author{Julian Schwarze}
% \affiliation{%
%   \institution{Gruner + Jahr GmbH}
%   \city{Hamburg}
%   \country{Germany}
% }
% \email{schwarze.julian@guj.de}

%%
%% By default, the full list of authors will be used in the page
%% headers. Often, this list is too long, and will overlap
%% other information printed in the page headers. This command allows
%% the author to define a more concise list
%% of authors' names for this purpose.
%\renewcommand{\shortauthors}{Wiese et al.}

%%
%% The abstract is a short summary of the work to be presented in the
%% article.

\maketitle

\begin{abstract}

Technical Debts (TD) are problems of the internal software quality. 
They are often contracted due to tight project deadlines, for example quick fixes and workarounds, and can make future changes more costly or impossible.
TD prevention should be more important than TD repayment, because subsequent refactoring and reengineering is usually more expensive than building the right solution from the beginning. 
While there are numerous works on TD repayment, solutions for TD prevention are understudied.
This paper evaluates a framework that focuses on both TD prevention and TD repayment. It was developed by and applied in an IT unit of a publishing house. 
The unique contribution of this framework is the integration of TD management into project management. 
The evaluation was carried out by a study based on ticket statistics and a structured survey with participants from the observed IT unit and a comparison unit.
The evaluation shows that the adoption of this framework leads to a raised awareness for the contraction of TD. 
This results in benefits like more rational discussions and decisions, TD prevention and timelier repayment of TD tickets.

\end{abstract}


%% A "teaser" image appears between the author and affiliation
%% information and the body of the document, and typically spans the
%% page.
% \begin{teaserfigure}
%   \includegraphics[width=\textwidth]{sampleteaser}
%   \caption{Seattle Mariners at Spring Training, 2010.}
%   \Description{Enjoying the baseball game from the third-base
%   seats. Ichiro Suzuki preparing to bat.}
%   \label{fig:teaser}
% \end{teaserfigure}

%%
%% This command processes the author and affiliation and title
%% information and builds the first part of the formatted document.

	\section{INTRODUCTION}
	\label{section:introduction}

% ## Begriffseinführung ## 
    In their definition of Technical Debts (TD), Avgeriou et al. \cite{Avgeriou2016a} describe TD as ``a collection of design or implementation constructs that are expedient in the short term, but set up a technical context that can make future changes more costly or impossible.''
    %``In software-intensive systems, technical debt is a collection of design or implementation constructs that are expedient in the short term, but set up a technical context that can make future changes more costly or impossible. Technical debt presents an actual or contingent liability whose impact is limited to internal system qualities, primarily maintainability and evolvability.''
	In a technical metaphor to financial debt a sub-optimal implementation or design is interpreted as debt. 
	The resulting problems are interpreted as interest rates, the refactoring as repayment of the debt and the refactoring cost as principal.
	
    %The consequences of TD during software evolution and maintenance are discussed in many papers (e.g. \cite{Besker2016, Avgeriou2016a, Martini2015a, Li2015}). 
	%Avgeriou et al. \cite{Avgeriou2016a} describe this as follows: ``\textit{While the conceptual roots of technical debt imply an idealized, deliberate decision-making process and rework strategy as needed, we now understand that technical debt is often incurred unintentionally and catches software developers by surprise}''.
	    
% ## Repayment vs. Prevention ## 
    %Furthermore, TD repayment is focused by many papers while TD prevention is worth more consideration but understudied so far \cite{Li2015}. 
	Many papers focus on TD repayment while the topic of TD prevention is understudied (\cite{Li2015}). %papers dealing with TD prevention are rare. 
	However, in a long-term view, it can be expected that it is cheaper to implement the optimal solution right from the beginning especially when considering possible interest payments (\cite{Yli-Huumo2016,Rios2018}). 
	Therefore, TD prevention is worth more consideration (\cite{Rios2018}).

% ## Causes allgemein (inkl. Timeline Problem) ## 
    The originally named cause for TD contraction is the problem of tight project deadlines as described in the paper of W. Cunningham (\cite{Cunningham1992}). 
    Other causes have been identified by different research papers (\cite{Verdecchia2020, Li2015,Avgeriou2016a,Martini2014}), e.g. bad design decisions, unavailability of a key person, neglected technical improvements, lack of education or parallel development.
	
% ## Timeline Problem ## 
    Nevertheless, tight timelines are the TD cause that was identified by many researches as the most pressing cause (\cite{ Freire2020, Rios2018, Martini2014, Verdecchia2020, Avgeriou2016a}). %,Buschmann2011,Malakuti2020
    However, to the best of the authors notice no studies provide a solution for the problem of tight timelines that is feasible in an industry environment.
	
% ## Beispiel (Timeline-bezug) ## 
	%Andere Formulierung fällt mir nicht ein:
    A good example of the timeline problems can be taken by the German government's reduction of value-added taxes from 07/01/2020 to 12/31/2020 as part of the management of the corona crisis. 
    The corresponding law was only passed two days before it came into force. 
    To reflect this change in the software systems, adjustments to these systems were necessary in many German IT units, which were therefore subject to a very tight schedule. 
    Well-structured project planning or adoption of good development practices were mostly not possible. 
 
% ## Developer Phenomenon ## 
    In many cases, the constant time pressure leads to a phenomenon in developer behavior. 
    In preemptive obedience the developers search for the fastest and easiest solution assuming a tight schedule, even when there is no time pressure. 
    In these cases, it is not common practice to step back and think about or discuss different solution options. 
    There may be frequent situations where the sub-optimal but faster to implement solution must be chosen due to project timelines.
    However, it should be a goal to make it a conscious decision of the whole team to choose the optimal or sub-optimal solution.
  
% ## TD tickets ##  
    At this point, TD prevention has a good chance for success by introducing a special form of tickets.
    These tickets that we call TD tickets make the contraction of TD explicit. 
    The awareness for the contraction of TD is raised among all team members including business analysts and managers.
    By this, a basis for TD prevention is created.
    
    Furthermore, TD tickets integrate TD management into project management.
    Project managers become responsible for the repayment of the TD they contracted during their project and therefore get an extrinsic motivation to reduce the contraction of TD.
    %ToDo: Hier Feedback-Loop ergänzen: Bei Project Manager wird extrinsisch der Bedarf erzeugt, TD zu verhindern. 
    By this, this paper presents a solution to the problem that is presented by Martini et al. (\cite{Martini2014}) as ``\textit{Split of budget in Project budget and Maintenance budget boosts the accumulation of debt}''. 

    The evaluation in this paper shows that TD tickets prevent TD and achieve a timely repayment of TD items.
    
% ## Framework##  
    These TD tickets are part of a framework developed in industry and a new contribution to the State of the Art. 
    Further parts of the framework are continuous TD repayment and maintenance projects which are already described in some research papers, e.g. by Steve McConnell (\cite{McConnell2008a}). 
    
% ## Contribution ##  
	The contribution of this paper is therefore the presentation and evaluation of the TD tickets as a means of TD prevention.
	The evaluation consists of a survey and ticket statistics. 
	It will provide evidence that TD prevention and timely TD repayment can be reached by adopting this framework. 
	Other benefits like more rational discussions and decision-making are highly appreciated side-effects.

% ## Outline ##   
	In the following section we will introduce the framework. 
	Section~\ref{section:rq} will present the research questions (RQs) regarding the effectiveness of the framework that will be answered by this paper. 
	The methodology of the evaluation will be explained in Section~\ref{section:method}. 
	The results of the evaluation will be presented in Section~\ref{section:results} and comprise ticket statistics and the results of a survey carried out in the observed IT unit and a comparison unit. 
	The results will be discussed in Section~\ref{section:discussion}. 
	In this section the RQs will be answered, threats to validity will be discussed and the paper will be embedded in related work. 
	The paper ends with the conclusion and future work in Section~\ref{section:conclusion}.

	\section  {FRAMEWORK PRESENTATION}
	\label{section:framework}
	
% Vorstellung des Untersuchungsgegenstands: Verhindern TD
% <erklärender Text>
% Challenges/Probleme
% - Verringerung von TD: durch Management
% - Verhinderung von TD: 
%     - durch Awareness, 
%     - durch Integration in Projektmanagement: durch Verteilung der Verantwortung
% Lösungen: Kategorien

	A framework for managing TD has been developed and established in an IT unit in the beginning of 2018. It has been utilized ever since. 
	The IT unit resides in a German publishing house with a size of more than 9000 employees worldwide.	
	
	In this IT unit, TD accumulation leads to increasing cycle times, unnecessary errors in the systems, and discontent of the developers. 
	The biggest problems of this IT unit are tight timelines. 
	There is a high market pressure in this industry sector and contracts or laws need to be fulfilled in time. 
	The tight timelines often mean that ``\textit{well-structured project planning}'' or ``\textit{adopting good development practices}'' as proposed by \cite{Freire2020} are not sufficient.

    These problems are recognized by the management and sufficient priority to handle TD is given. 
    The IT unit including the management is willing to change its processes.
    The main goal for this change is predefined by the manager of the IT unit: TD have to be prevented in any case where this is possible without impairing the deadlines. Whenever a set deadline leads to TD, this TD shall be repaid timely.
    
    On this basis, a framework to help manage TD was created. 
    Especially, two factors of this framework shall support TD prevention: the overall raised awareness for TD contraction and the integration of TD management in the project management process.
	The resulting framework consists of four management categories for TD management which are presented below.

		\subsection  {Management Categories}
		\label{section:ManagementCategories}
	
	    The framework comprises four management categories for handling different kinds of TD related tasks.
	    There is a fifth management category for handling all functional requirements which will not be presented in detail as it does not deal with TD. 
	    
	    The different tasks are recorded as tickets in a comprehensive project backlog and tagged and handled depending on their category.
		The framework includes guidelines for the recording and processing of these tickets depending on their management category. 
		It was decided that the person that has the most interest in the completion of the tasks of one category should be the one that is responsible for this category. These may be architects or business analysts. 
	    Table~\ref{tab:solution} shows an overview of the management categories, their responsible persons and repayment terms.

		\begin{table*} 
		    \centering
			\caption{Ticket categories, responsible persons and time}
			\label{tab:solution}
			\begin{tabular}{llllll}
				\toprule
				Management Category		& Responsible		& Recorded By		& Payback Time						& Contingent\\ 
				\midrule 
				Maintenance  		 	& architect	 		& developer			& continually / architect decision 	& maintenance (10\%) \\ 
	 			Maintenance Project 	& architect 		& architect			& management decision				& project\\ 
				Technical Debt  	 	& business analyst	& all 				& after deadline, part of project	& project \\
	 			Deconstruction 	 	    & architect 		& all				& as soon as possible				& project / functional req.\\ 
				Functional Requirement  & business analyst & business analyst & business analyst decision			& project / functional req.\\ 
	 			\bottomrule
			\end{tabular}
		\end{table*}

		\subsubsection {Maintenance}
		In this framework, tickets of the maintenance category are technically driven tickets that do not have a direct impact on the user's perception of the system. 
		%Adaptive maintenance, e.g. adaption to new technical requirements as well as repayment of unintentionally contracted TD, are the main tasks covered by this category. 
		%Corrective maintenance is only included in this category if it is ``invisible'' to the user, which follows the distinction also made in \cite{Kruchten2012a}. %The category of corrective maintenance tickets is therefore assigned according to the visibility. 
    	%The label "maintenance" is based on the term like it is often used in agile organizations and adaption in terms of new functional requirements is not considered a part of maintenance in this IT unit. 
        \footnote{We use the term ``maintenance'' as it is often used in agile organizations. However, implementation of new or changing functional requirements is called adaptive maintenance in research, standardization, and sometimes in classical organizations.}% (\cite{ISO/IEC2006a}). }
    	%, but is the core of its agile development. It is therefore not included in this category. On the other hand, 
    	
		\textit{Recording:} Every maintenance ticket must contain a one-sentence information that describes the impact of the ticket. 
		The description must be written in a way that is understandable for business analysts and managers. 
		\textit{Processing:} Ten percent of the planned capacity of every sprint can be invested for maintenance tickets according to the architect's decision. 
		%Streichen?%
		%This follows a common idea of continual TD repayment, e.g. as proposed in \cite{McConnell2008a}.
		\textit{Goal:} The goal of the maintenance tickets is to repay the unintentionally contracted TD continually and to allocate time for other maintenance tasks, e.g. technical adaption.

	    Examples for this category are the upgrade of a third-party-library or the refactoring of bad structured code. %Streichen?
		%detected by a static analysis tool or by chance while implementing something else.

		\subsubsection  {Maintenance Project}
		All tasks that belong to the maintenance category as stated above but that require more than five days development time are handled as a maintenance project. 
		This is based on the practice that is also applied for functional requirements and business projects. %where small functional adaptions are part of continual work and bigger requirements will be implemented in the context of projects.

		\textit{Recording:} %Streichen?
		%Maintenance projects are handled like a business project. 
		%Therefore, 
		A maintenance project roadmap is produced and prioritized by the architects. 
		\textit{Processing:} The roadmap is part of the overall project roadmap as defined and prioritized by the unit managers. 
		\textit{Goal:} The goal of maintenance projects is to allocate enough time to conduct maintenance tasks with a long duration or maintenance tasks requiring more organization, e.g. when more than one team is involved. 

		An example for this category is the version upgrade of a central database that is used by more than one application. 
		Another example would be the change of the underlying architecture, e.g. from a SOA to a microservice architecture. 
		%Streichen?
		%Both will take a lot of time and needs the collaboration of different teams as more than one application is affected. 
		%This in turn makes a project planning necessary.

		\subsubsection  {Technical Debt}
		Tickets of this category, so-called TD tickets, describe tasks that are necessary for internal software quality during the implementation of a functional requirement, e.g. time for clean coding, good design or to comply with a specified architecture. 
		%Streichen?
		These tasks are not mandatory to implement the required functionality and can therefore be implemented after a reached deadline if this is necessary for the project.
		
		Often two or more solutions are discussed, where one follows standards and the specified architecture and the others more or less deviate from the optimal solution but may therefore be faster to implement. 
		This idea of TD follows its original introduction by Cunningham (\cite{Cunningham1992}). 

\begin{figure} % [ht]
	 		\centering
			\includegraphics[width=\linewidth]{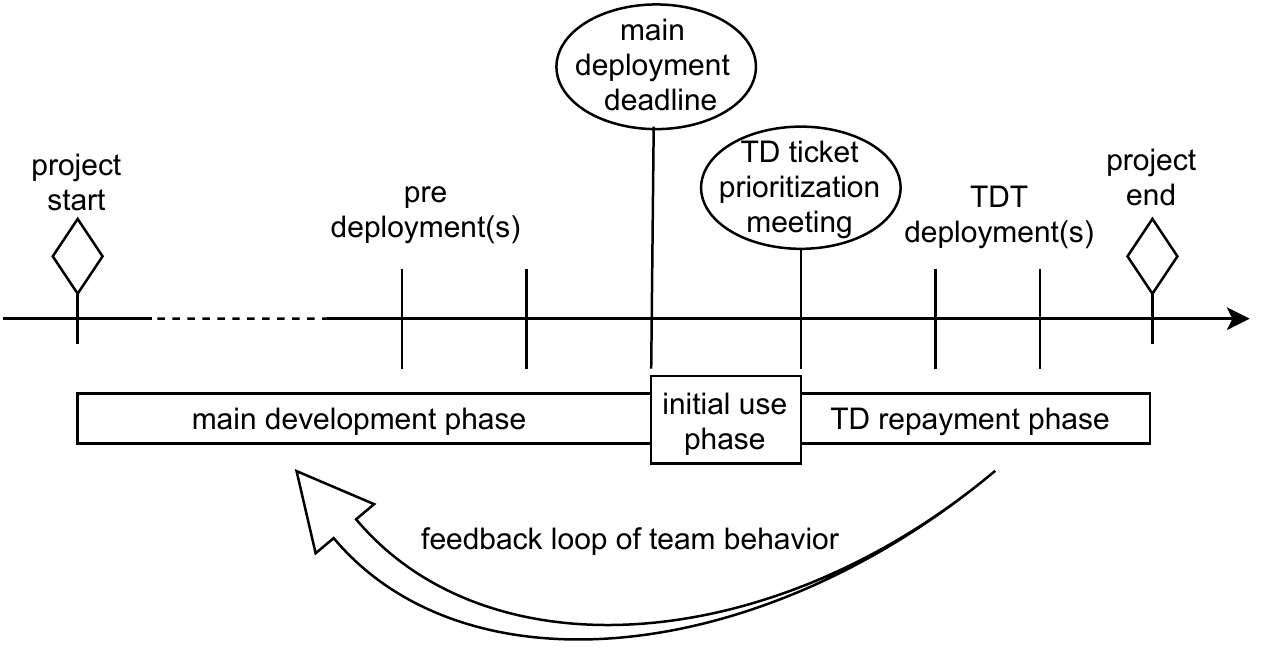}
			\caption{Project plan with included TD repayment phase}
	 		\label{fig:ProjPlan}
  		%	\Description{After the main deployment a phase of TD repayment follows.}
		\end{figure}

		\textit{Recording:} In such cases, it is consciously decided to record two tickets. 
		One ticket, a functional requirement ticket, describes the sub-optimal solution to be implemented before the deadline. By this ticket, TD is contracted. 
		The other ticket, the TD ticket, describes the optimal solution and the tasks to repay the contracted TD. 
		These TD tickets are usually identified during estimation meetings when the details of a functional requirement are discussed in the team and the effort is estimated. 
		%Streichen?
		%This is also why, most of the TD tickets span mainly design and architectural TD and not code or other TD, e.g. unintentional TD.
		\textit{Processing:} The uniqueness of this framework is that the repayment of intentionally contracted TD is part of the project plan. 
		Thus, no project can be finished before the TD tickets are processed. 
		After reaching the planned milestone for the main deployment and keeping to the tight schedule, the project plan does not end. 
		Instead, TD tickets are prioritized and repaid in a subsequent project phase. 
		This can be seen in Figure~\ref{fig:ProjPlan}. 
		This transfers the responsibility for the TD repayment to project management and thereby integrates TD management into project management. 
		\textit{Goal:} %One goal of TD Tickets is to raise the overall awareness to only take on TD intentionally.
		Evaluating different solution options in advance raises the overall awareness to only take on TD intentionally and prevents unnecessary TD. 
		Furthermore, by making the project manager responsible for the TD accumulated during the project, his willingness to contract TD decreases.
		In other words, a feedback loop is generated and the behavior of the team and the project manager related to TD contraction changes.
		Additionally, the project and unit managers get an overview of the accumulating technical debt of a given project while it is still running.
		Thereby, the managers are given an opportunity to intervene or adjust the project plan and the following projects early on. 
		Finally, by this approach TD are repaid timely.

		For the example from Section~\ref{section:introduction} this may mean that, to adhere to the schedule, hard coded value-added taxes are adjusted in the code and are not refactored to a flexible solution. 
		The corresponding TD ticket to introduce a centralized tax variable, is recorded during the estimation meeting. 
		This TD will then be repaid in an orderly manner after the law came into force and in the ``TD repayment phase'' of the ``tax adjustment'' project. 
		On the other hand, as different options for these changes are discussed during the estimation meeting, this can also mean that due to the low effort the centralization may be the better option and TD are prevented.

		\subsubsection  {Deconstruction}
		\label{section:deconstructionTickets}
		A ticket of the deconstruction category is a special kind of TD ticket that cannot be processed during the project. 
		After the implementation of a new solution there is sometimes the need to keep legacy code in parallel for the time being.
		These tickets then comprise the deconstruction of this legacy code when it is no longer needed. 
		They can also comprise the deletion of deprecated code parts after all consumers of this parts are detached. 
	
		\textit{Recording:} These tickets have to contain an additional information describing the time when the deconstruction can take place. 
		\textit{Processing:} The deconstruction ticket has to be repaid as soon as possible as part of the contingent for functional requirements.
		\textit{Goal:} The main goal of this tickets is to avoid cluttering of code and by this support code comprehension. 
		
		An example of this kind of tickets is the need for a new business rule for the next business year. 
		The application should not be deployed on New Year's Eve to minimize deployment risks. 
		Therefore, the application will be deployed in advance and must contain a date-driven switch to choose between the old and new business rule. 
		When the old business rule is no longer needed it can be deconstructed. 
		%Streichen?
		%The ``point in time'' in this ticket would be January, 1 of the respective next year.

	\section{RESEARCH QUESTIONS}
	\label{section:rq}
	As described in Section~\ref{section:introduction}, two main problems concerning TD are the questions on how to prevent TD and how to deal with TD despite having a tight schedule and fixed deadlines. 
	For this question, the described framework provides a possible solution found and developed in practice. 
	With subsequent research questions (RQ) this paper provides an evaluation of this framework. Especially its feasibility and its effectiveness regarding the aforementioned problems is evaluated.
	
    \textit{(\textbf{RQ1}) Framework application}
    \begin{itemize}
        \item \textit{(\textbf{RQ1.1}) Do practitioners find the framework reasonable?} 
        \item \textit{(\textbf{RQ1.2}) Are the processes of the framework feasible in practice?}
    \end{itemize}
	This will be the basis for all other evaluations as it does not make sense to further analyze the impact of the framework should it not be feasible in practice. 
	These questions will be evaluated by a survey of the members of the observed IT unit.
	
    \textit{(\textbf{RQ2}) Framework effectiveness}
    \begin{itemize}
        \item \textit{(\textbf{RQ2.1}) Does the framework lead to raised awareness for the contraction of TD?}
        \item \textit{(\textbf{RQ2.2}) Are the TD items taken on more consciously when using the framework?}
        \item \textit{(\textbf{RQ2.3}) Are the TD items paid back timely?}
    \end{itemize}
	These aspects are the goals of the framework as presented in Section \ref{section:framework}. 
	The evaluation of this question shall show whether these goals were reached. 
	The evaluation is conducted by a survey and the tickets statistics (for the last part). 
	
    \textit{(\textbf{RQ3}) Framework benefits}
    \begin{itemize}
        \item \textit{(\textbf{RQ3.1}) Can TD be prevented by the adoption of the framework?} 
        \item \textit{(\textbf{RQ3.2}) Are there other benefits arising from the adoption of the framework?}
        \item \textit{(\textbf{RQ3.3}) Do these benefits justify the additional effort?}
    \end{itemize}
	The purpose of these questions is to gather the impact of the framework. 
% 	The main goal of TD prevention is evaluated.
% 	Secondary benefits concerning team discussion and decision-making process will be evaluated, also. 
	The main goal of TD prevention is evaluated, as well as secondary benefits concerning team discussion and decision-making process. 
	The correlation between the survey variables will be used to answer this RQ.
	Finally, we ask for the justification of the additional effort in relation to the benefits to substantiate the feasibility of the framework.

	\section{RESEARCH METHOD}
	\label{section:method}
	The presentation and evaluation of the aforementioned framework is the contribution of this paper.
	The presentation took place in Section \ref{section:framework}. 
	To answer RQ 1.2 and RQ 2.2, we evaluated the recording and processing statistics for TD tickets and maintenance tickets.  
	All other RQs are answered by the results of a survey. 
	The survey targets participants from the IT unit that uses the framework (observed IT unit) and a comparison unit that does not use the framework.

	\subsection{Ticket Statistics}
	\label{section:TicketStatistics}
	To evaluate RQ 1.2 and RQ 2.2 the number of tickets are presented in terms of categories, priorities, and time as descriptive statistics. 
 	As the backlog includes some old and invalid tickets the extracted data had to be cleaned up manually first. 
 	This was carried out in consultation with the architects of the unit. 
	The statistics for maintenance projects and deconstruction tickets are not evaluated because there is still insufficient data for this. 

	\subsection{Survey}
	\label{section:SurveyMethod}
	The method of a survey was chosen to benefit from the input of all participants that are willing to share their experience. 
	For all RQ's corresponding questions were asked in this questionnaire.
	
	\subsubsection{Survey Participants}
	\label{section:SurveyParticipants}
	All members of the observed IT unit are asked to fill out the questionnaire. 
	Additionally, a comparison unit is asked to complete the questionnaire. 
	The comparison unit is led by the same unit manager which increases the comparability.
	
	The results of the comparison unit gave us the opportunity to validate the descriptive statistics.
	The participants were asked to which team they belong, but due to works council regulations not which role (e.g. architect, manager) they inhibited.
	The composition of the observed unit and the comparison unit are presented in Table~\ref{tab:unitcomp}. 
	The response to the survey is also shown in this table. 
	All questions and evaluations of the survey can be accessed in the additional material\footnote{ \url{https://doi.org/10.5281/zenodo.4616485}}.

	\begin{table}
	    \centering
		\caption{Composition of units and survey participants}
		\label{tab:unitcomp}
		\begin{tabular}{lcccc}
			\toprule
				&\multicolumn{2}{c}{observed unit} 	&\multicolumn{2}{c}{comparison unit}\\ 
							&members&particip. 	&members&particip.\\ 
			\midrule 
			manager 			& 2& n/a 			& 2 	& n/a \\ 
			architect		 	& 2 & n/a 			& 1$^{\mathrm{a}}$    & n/a \\ 
			business analyst	& 8 & 6   			& 1$^{\mathrm{a}}$ 	& 1 \\
 			developer 			& 15& 9   			& 5 	& 4 \\ 
			operations  		& 5 & 2   			& 0 	& 0\\ 
			\midrule 
			sum  				& 32& 17  			& 8 	& 5\\ 
 			\bottomrule
 			\multicolumn{5}{l}{$^{\mathrm{a}}$architect and business analyst are the same person}
		\end{tabular}
	\end{table}

	\subsubsection{Questionnaire Construction}
	\label{section:QuestionnaireConstruction}
	To avoid misunderstandings the questionnaire starts with a short information about the term TD. 
	For statistical reasons, background information, e.g. the team membership of the participant, are queried.
	The questionnaire is further divided into two main parts: (I) assessment of the framework and (II) effects and benefits of the framework. 
	
	Part (I) shows the practical feasibility of the framework and provides the answers to RQ1.1 and RQ1.2. 
	For every management category of the framework the same set questions regarding the reasonableness and feasibility are asked. 
	Each part starts with a short information about the management category as it was also documented in the IT unit's wiki.
	Part (I) was not filled out by the comparison unit as these participants did not use the framework. 

	The second part comprises four subsections: TD Awareness (RQ2.1), Comparison of optimal and sub-optimal solutions (RQ 2.2), Observed benefits of comparison (RQ3.1 and RQ3.2), and Justification of the additional effort (RQ 3.3).
% 	\begin{itemize}
% 	    \item TD Awareness (RQ2.1)
% 	    \item Comparison of optimal and sub-optimal solutions (RQ 2.2)
% 	    \item Observed benefits of comparison (RQ3.1 and RQ3.2)
% 	    \item Justification of the additional effort (RQ 3.3)
% 	\end{itemize}
	
	All parts comprised a set of assertions that the participants were asked to validate using the following Likert scale (\cite{Joshi2015}): applies - rather applies -  rather does not apply - does not apply - cannot answer. 
	The \textit{cannot answer} option was given as some questions could not be answered by all members of the unit.
    The participants were explicitly asked to only use this option in these cases. 
	
	Finally, two more open questions in the end gave the participants an opportunity to point out especially good parts of the framework and parts that need improvement.

	A first questionnaire was filled by two developers that knew about the framework as a pilot test. 
	Based on this, %the questionnaire was reworked, and 
	the questions were reduced and focused. 
	Furthermore, one manager gave feedback on the construction of the optimized questionnaire which was also incorporated.

	\subsubsection{Survey Evaluation}
	\label{section:SurveyEvaluation}
	
	The questionnaire is evaluated using the statistics software SPSS\footnote{\url{https://www.ibm.com/dede/analytics/spss-statistics-software}} for closed questions and the qualitative research software MaxQDA\footnote{\url{https://www.maxqda.de/}} for open questions.
	
	First, descriptive statistics for all closed questions are created for an exploratory data analysis.
	For simplification and analysis purposes the answers for the questions of part (II) are dichotomized. 
	This means the answers \textit{applies} and \textit{rather applies} as well as \textit{rather does not apply} and \textit{does not apply} are summarized. 
	All values are presented as percentage to align the output of the observed unit and the smaller comparison unit. 
	For the sake of clarity, only the \textit{applies} answers are shown in the following figures. 
	
	As a second step, hypothesis for RQ2 are developed on the basis of this exploration and the significances of the hypotheses are evaluated. 
    The Mann-Whitney U-Test (\cite{Mann1947}) is used because it can be applied for not normally distributed data on an ordinal scale.
	Only significant findings are presented.
	
	Regarding the comparison unit, the benefits (RQ 3.2 and RQ3.3.) cannot be assessed but must be interpreted as expected benefits. 
	As a result, no differences between the units can be evaluated regarding the benefits. 
	Hence, it is interesting to analyze correlations between all survey participants who take on TD consciously and the benefits they assessed. 
	These correlations are evaluated using the $\phi$-coefficient of Pearson's $\tilde{\chi}^2$-correlation (\cite{Cohen1988,Pearson1900}) for dichotomous variables.
	Significant and strong correlations are deduced and presented.
	
	Lastly, the open questions were evaluated using open coding.

	\section{RESULTS}
	\label{section:results}

		\subsection  {Ticket Statistics}
		\label{subsection:TicketStat}

\begin{figure}
		 		\centering
				\label{fig:TicketbyTimePrio}
	 			\begin{tabular}{@{}c@{}}
					\subfigure[Maintenance tickets by time]
					{	\label{fig:MTbyTime} 
						\includegraphics[width=0.45\textwidth]{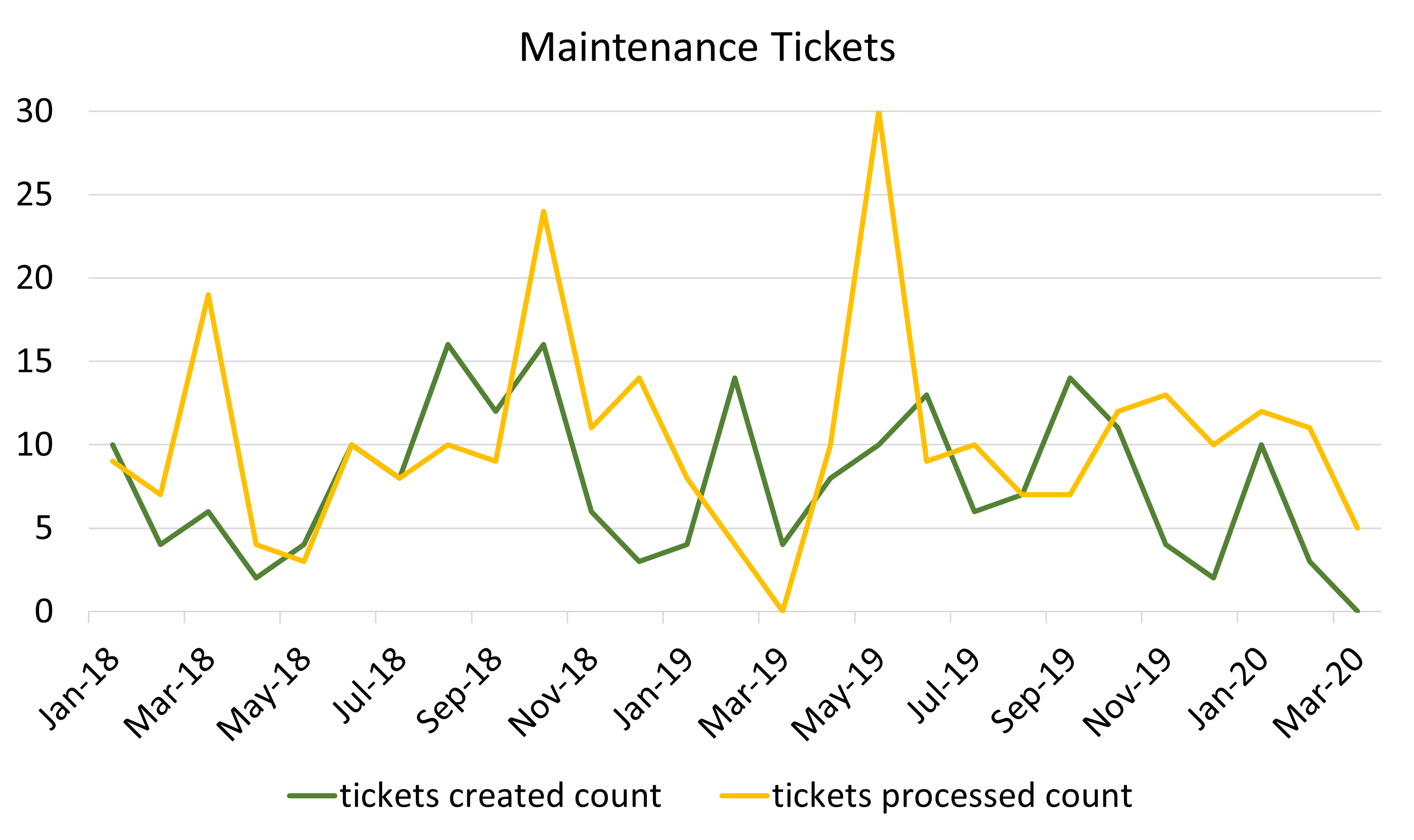}
					}\\
					\subfigure[TD tickets by time]
					{	\label{fig:TDTbyTime}
						\includegraphics[width=0.45\textwidth]{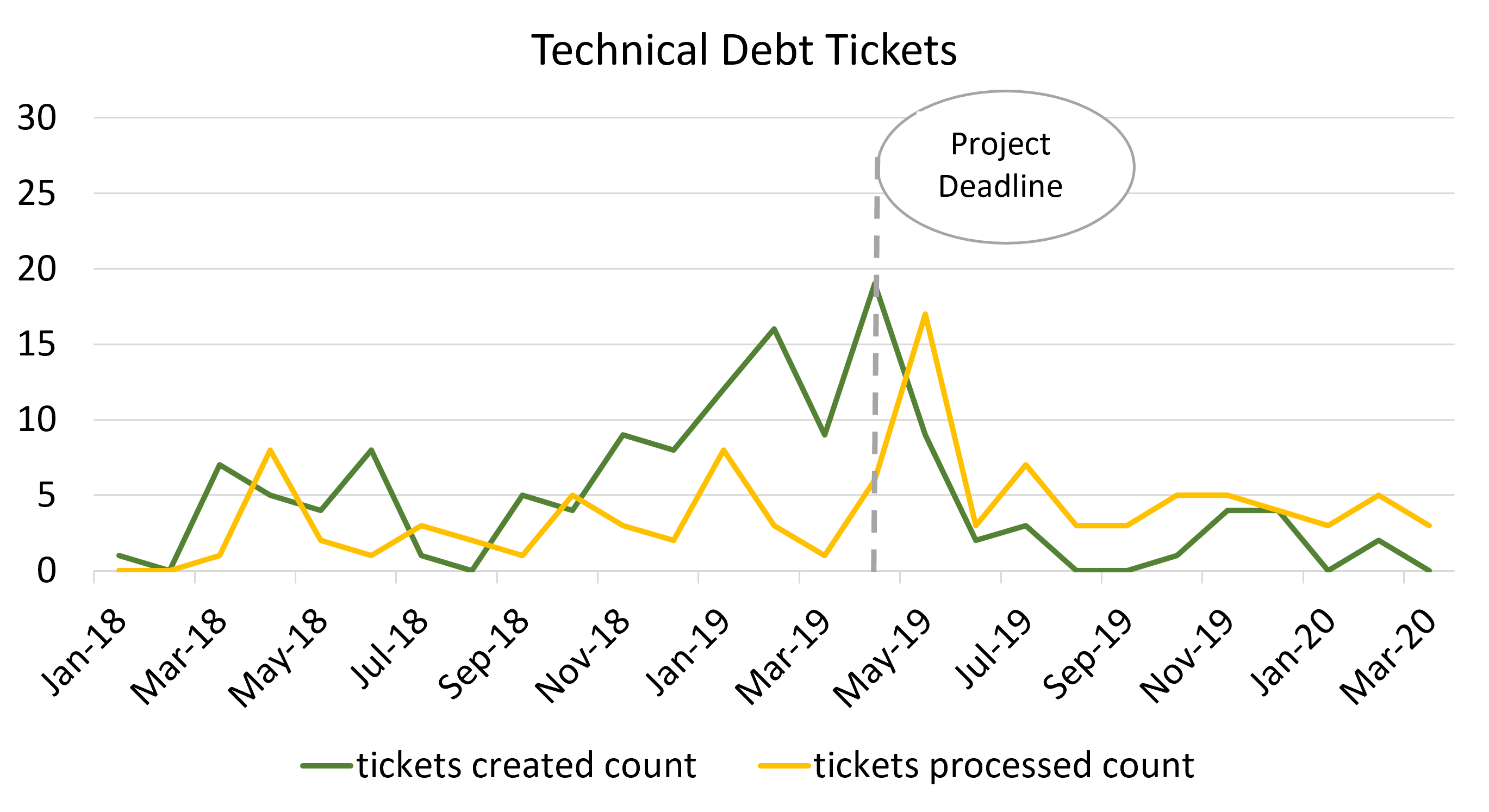}
					}
	 			\end{tabular}
				\caption{Ticket statistics with project deadline at the end of April 2019}
	  			%\Description{Maintenance tickets are constantly created and done. TD tickets are accumulated in spring and repaid in summer and autumn.}
			\end{figure}

		The timelines in Figure~\ref{fig:MTbyTime} and Figure~\ref{fig:TDTbyTime} show the development of ticket counts over time to answer RQ2.3. 
		The maintenance tickets are created and processed more or less continually.
		
		The timeline for TD tickets show a peak of contracted TD in spring 2019 due to a project with a tight timeline. 
		No maintenance or TD tickets were processed in March 2019. 
		Accordingly, the TD tickets as well as the maintenance tickets show a processing peak after the reached deadline at the end of April 2019. 
		In summer and autumn 2019 the TD tickets with a lower priority are processed which means that this project is still not finished (see Figure \ref{fig:ProjPlan}). 
		Other projects start in parallel to this project and development capacities must be divided between the projects.

		\subsection  {Survey}
		\label{subsection:survey}
 %---------------------------------------------------------------------------------
			\subsubsection  {Assessment of the Framework} 
			
\begin{figure*}
		 		\centering
				\label{fig:FwAss}

	 			\begin{tabular}{@{}ccc@{}}
					\subfigure[Is the recording of the ticket reasonable in general?]
					{	\label{fig:RecordingRes} 
						\includegraphics[width=0.31\textwidth]{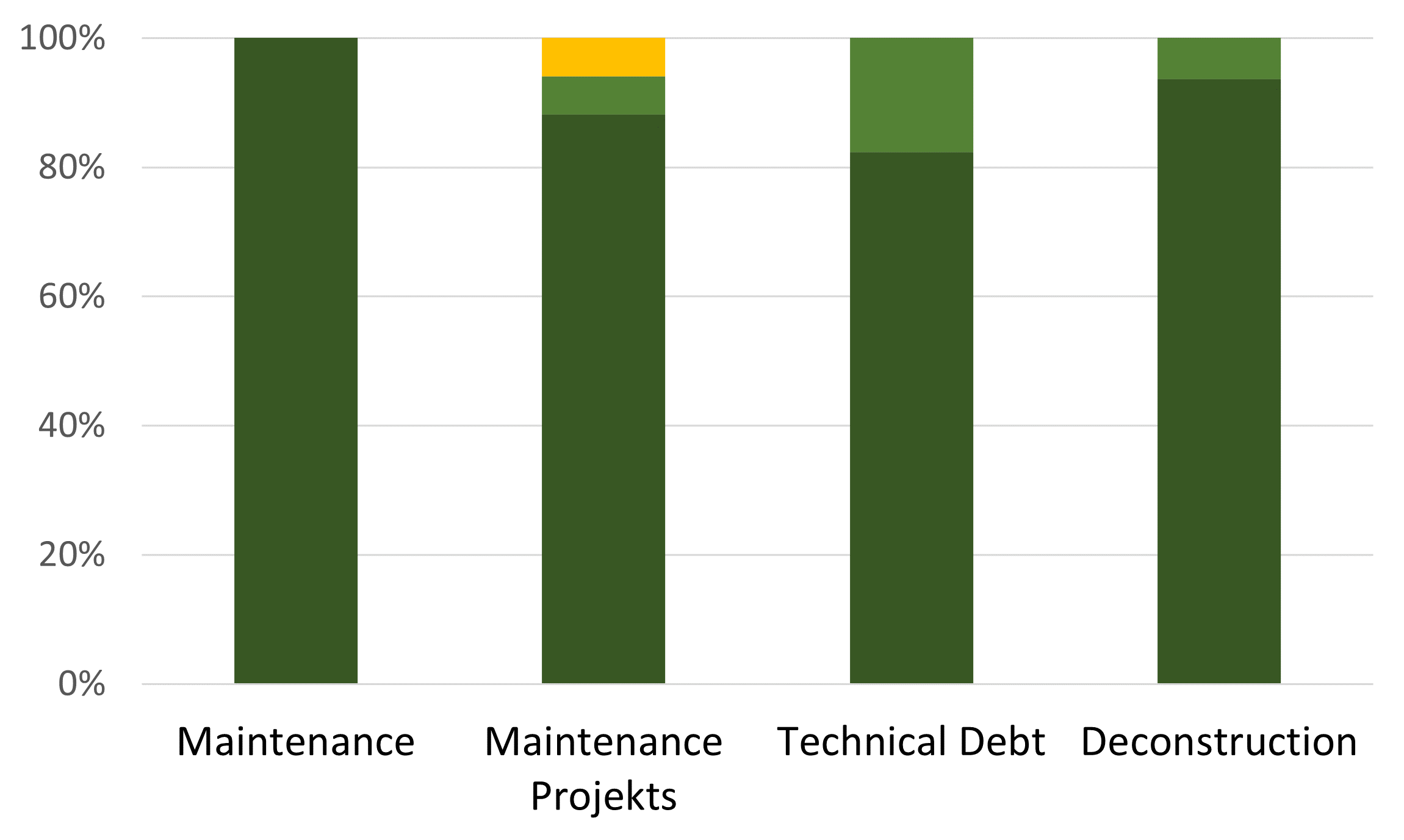}
					}&
					\subfigure[Is the procedure for recording the ticket reasonable?]
					{	\label{fig:RecordingProcRes} 
						\includegraphics[width=0.31\textwidth]{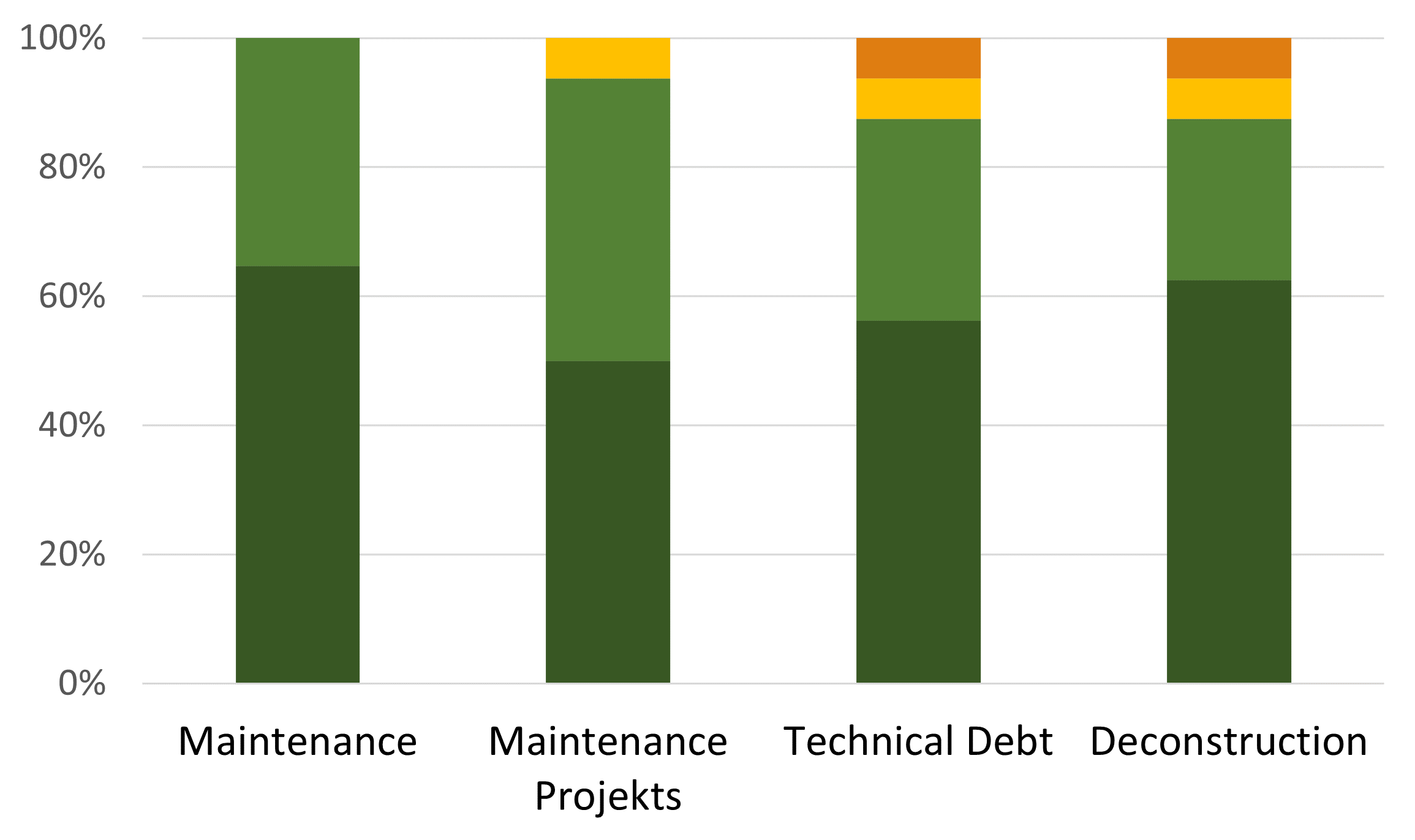}
					}&
					\subfigure[Does the procedure for recording the tickets work?]
					{	\label{fig:RecordingProcWork} 
						\includegraphics[width=0.31\textwidth]{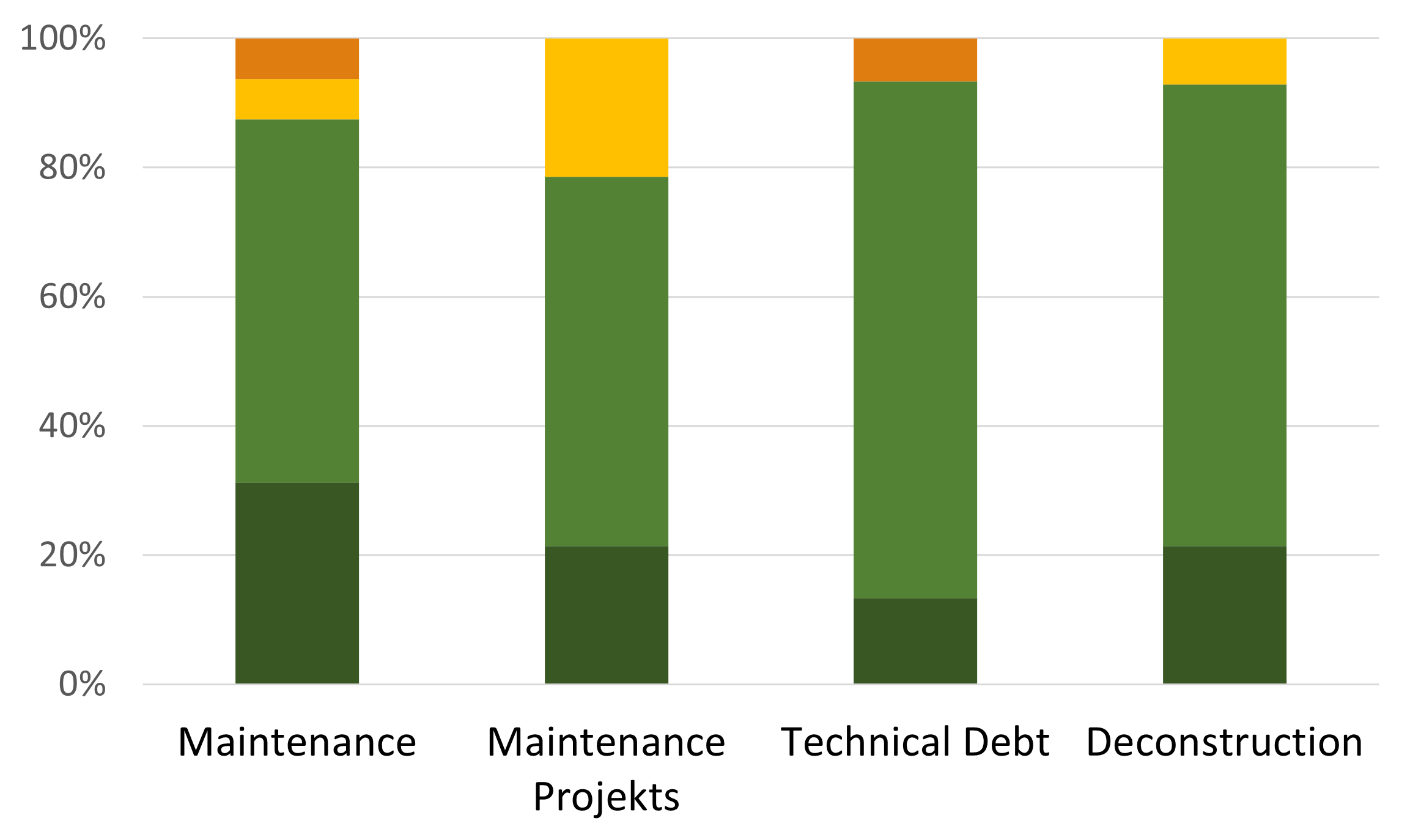}
					} \\
					\subfigure[Is the processing of the ticket reasonable in general?]
					{	\label{fig:ProcessingRes} 
						\includegraphics[width=0.31\textwidth]{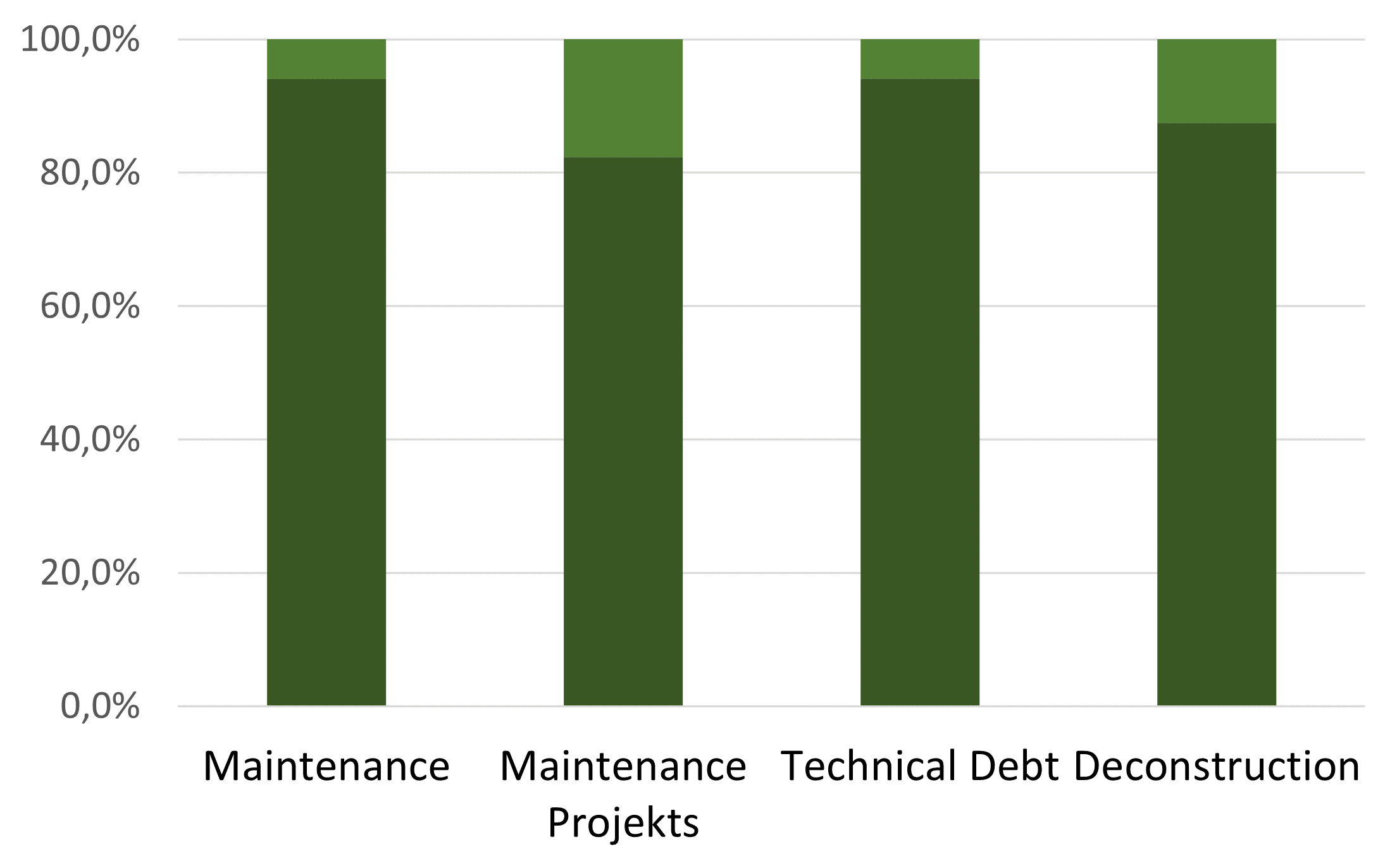}
					}&
					\subfigure[Is the procedure for processing the ticket reasonable?]
					{	\label{fig:ProcessingProcRes} 
						\includegraphics[width=0.31\textwidth]{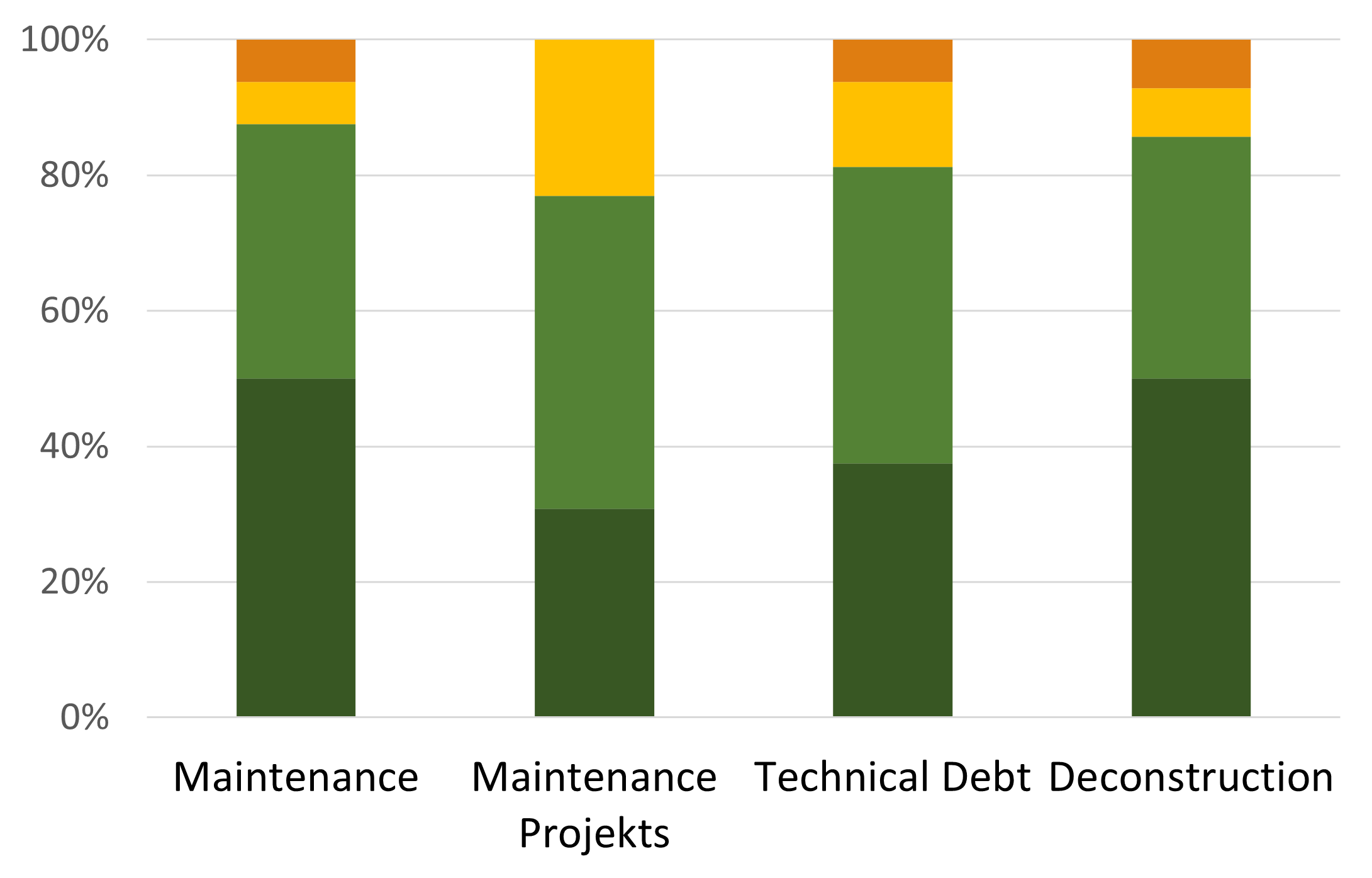}
					}&
					\subfigure[Does the procedure for processing the tickets work?]
					{	\label{fig:ProcessingProcWork} 
						\includegraphics[width=0.31\textwidth]{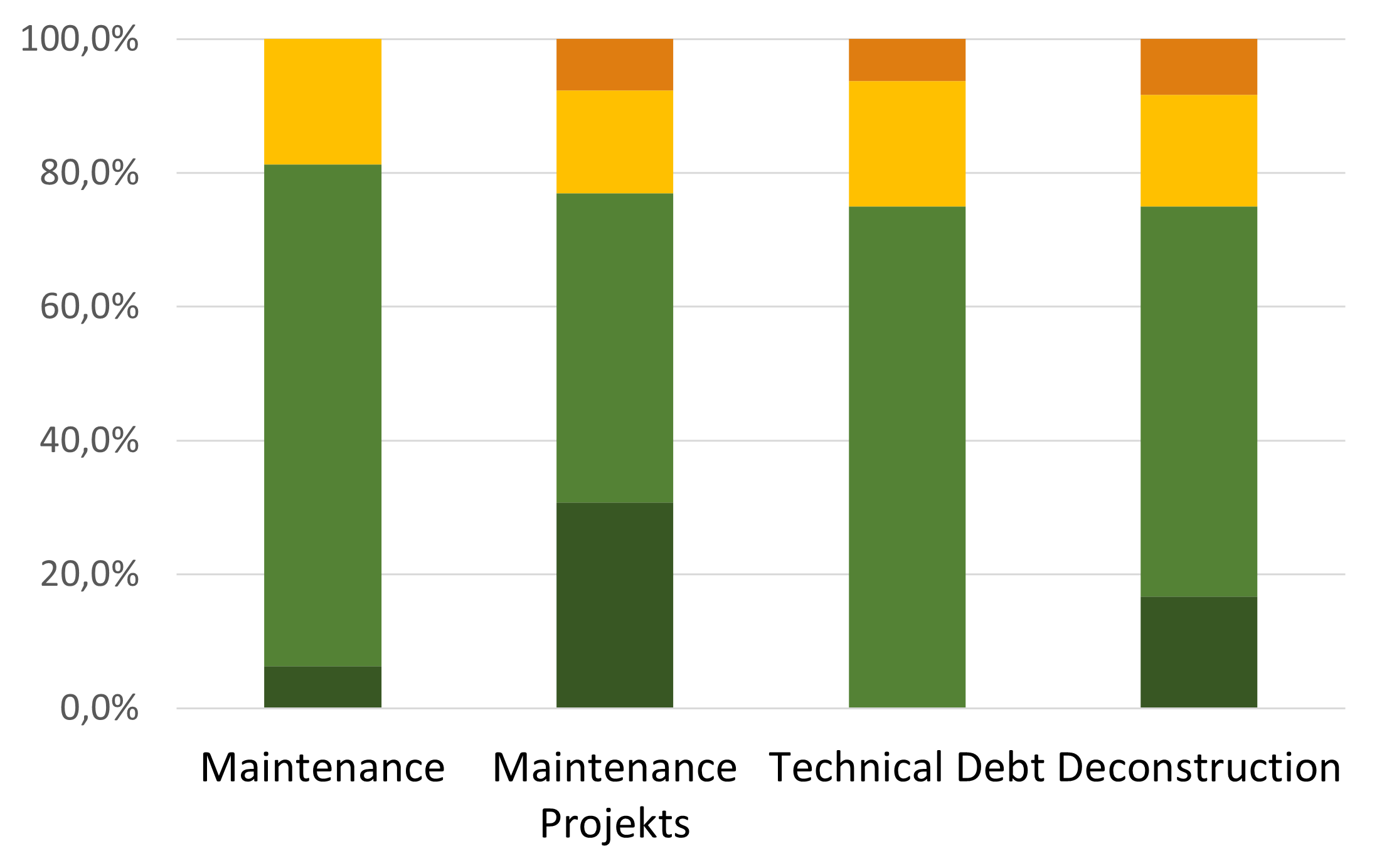}
					}\\
 				
					\multicolumn{3}{c}{\includegraphics[width=0.4\textwidth]{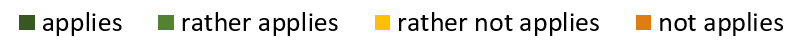} } \\
	
	 			\end{tabular}

				\caption{Appropriateness and usability of the framework as assessed by the observed IT unit}
	  			%\Description{Recording and processing of the tickets is reasonable, the chosen procedures are reasonable. The procedures work, but agreement is less.}
			\end{figure*}

			Most of the survey participants agree that it is generally reasonable to record and process all four types of tickets (Fig.~\ref{fig:RecordingRes}/~\ref{fig:ProcessingRes}). 
			Most of them also agree that the procedures of the framework for recording and processing are reasonable (Fig.~\ref{fig:RecordingProcRes}/~\ref{fig:ProcessingProcRes}). 
			When asked if the framework is working as intended, the agreement decreases, but still, most of the survey participants state that the procedures work well and are therefore helpful (Fig.~\ref{fig:RecordingProcWork}/~\ref{fig:ProcessingProcWork}).
			For RQ1.1 and RQ1.2 this means that the framework is reasonable and feasible in practice.

 %---------------------------------------------------------------------------------
			\subsubsection  {Effects and benefits of the Framework}  

			\begin{figure*} 
			\centering
			\begin{tabular}{@{}cc@{}}
				\subfigure[Do the survey participants recognize that TDs are taken on while contracting them?]
					{	\label{fig:RecTD} 
						\includegraphics[width=0.48\textwidth]{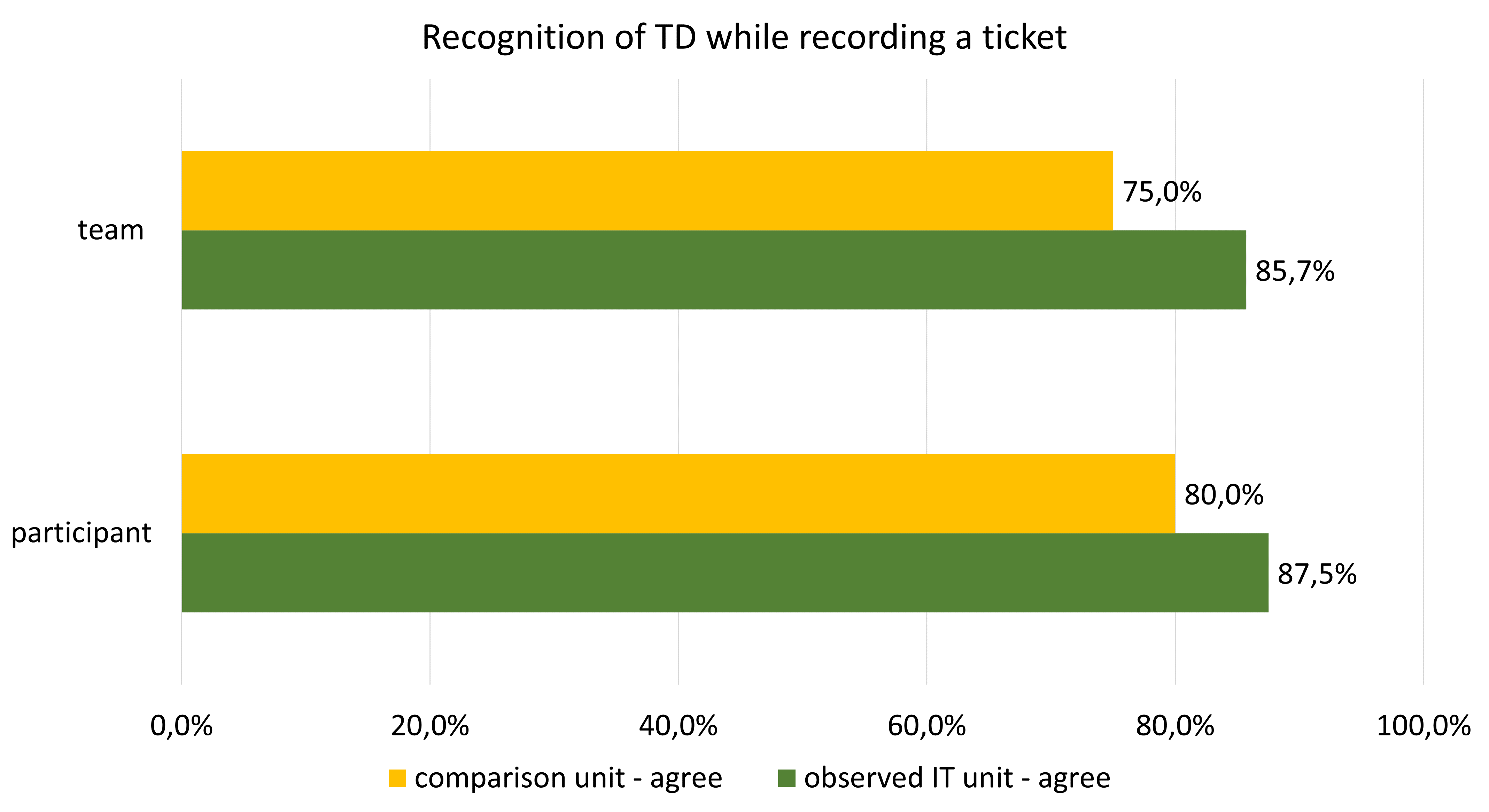}
			  			%\Description{No difference between the two units}
					}&
				\subfigure[To what extend are the quick/sub-optimal and the optimal solution compared?]
					{	\label{fig:CompareQvO}
						\includegraphics[width=0.48\linewidth]{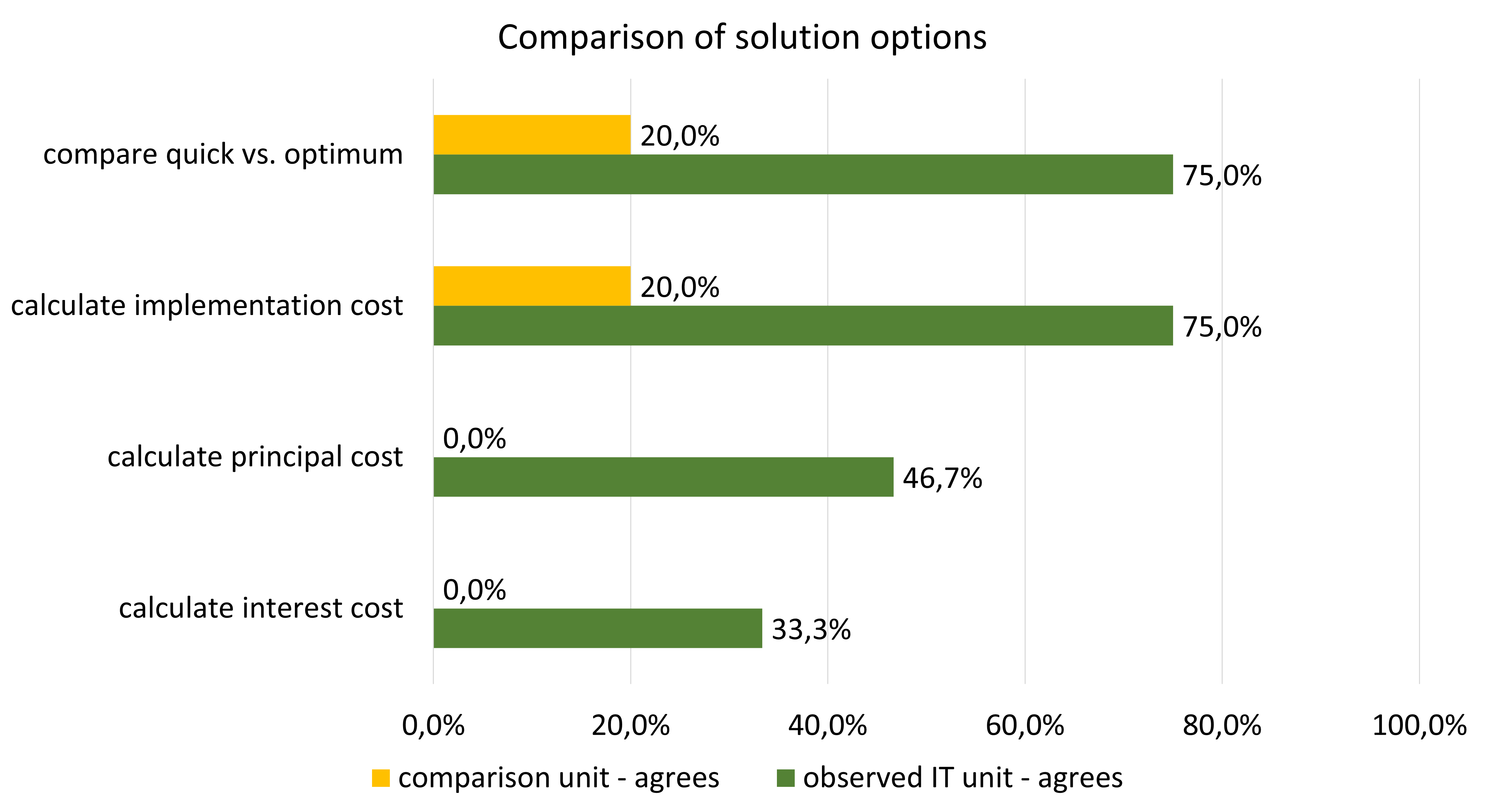}
			  			%\Description{A significant difference can be seen.}
					}\\
				\subfigure[What benefits of the comparison can be found?]
					{	\label{fig:benefits}
						\includegraphics[width=0.48\textwidth]{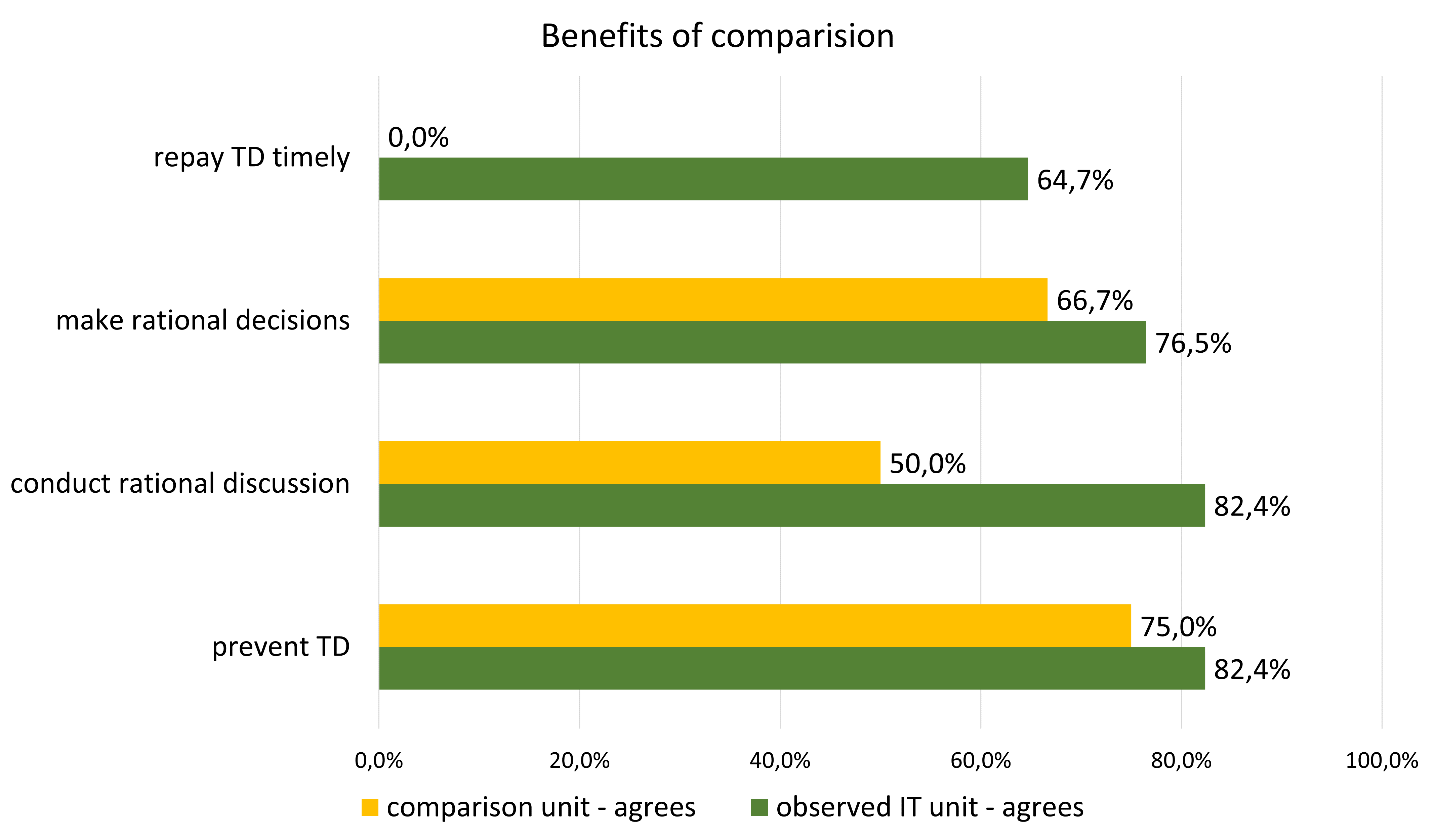}
			  			%\Description{A difference between groups can be seen, but that difference is not significant.}
					}&
				\subfigure[Do the survey participants think that the extra-effort for the comparison is justified?]
					{	\label{fig:EffortJustified}
						\includegraphics[width=0.48\textwidth]{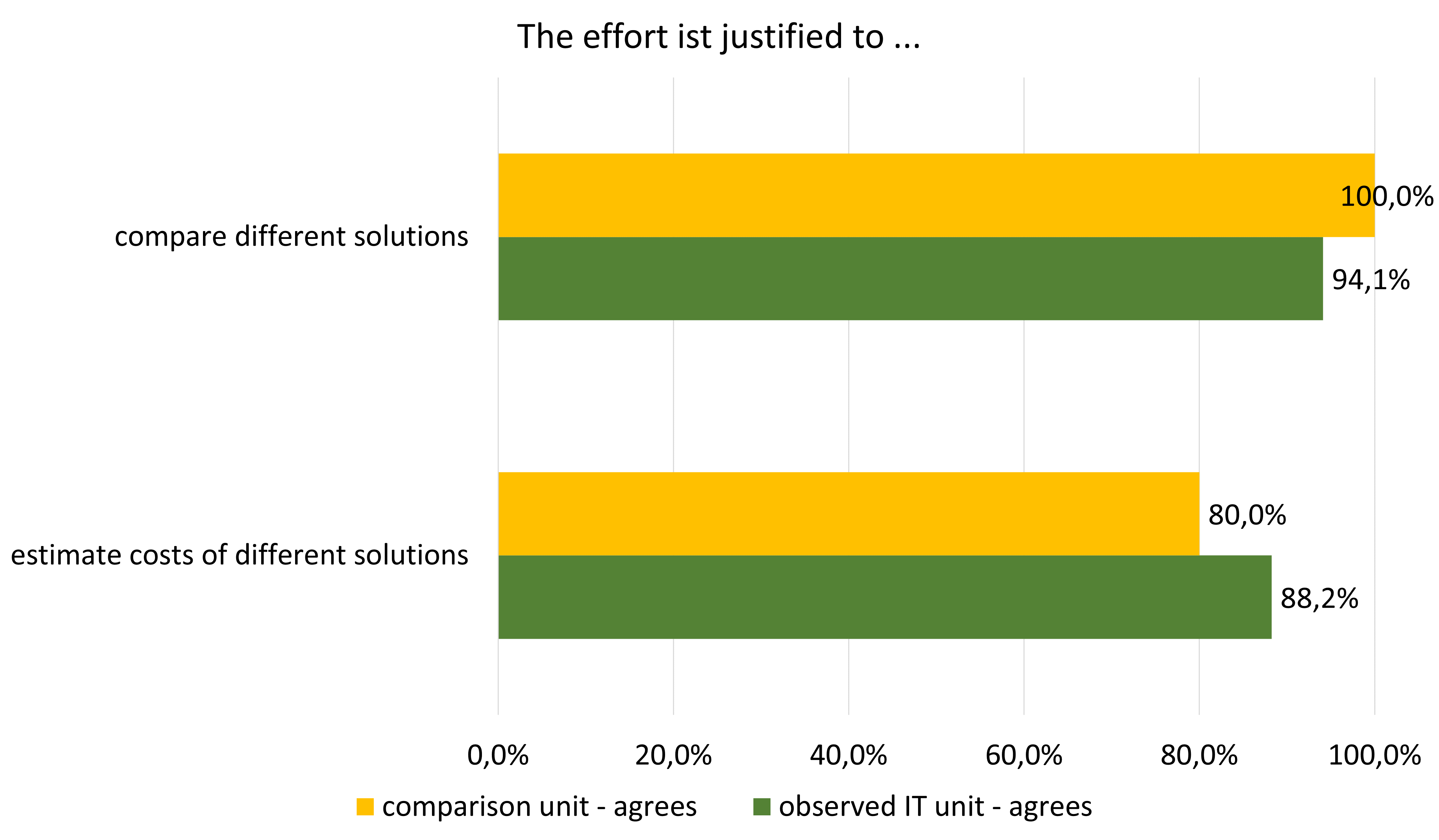}
			  			%\Description{All units find the extra effort to be justified.}
					}
	 		\end{tabular}

			\caption{Effects and benefits of the Framework - Descriptive Statistics}
			\end{figure*}
			
			First, the survey participants are asked if they and their team usually recognize that TD items are or were contracted before or after the point in time when they were contracted. 
			The goal of this question is to identify whether the participants are aware of the contraction of TD items to answer RQ2.1.
			As can be seen in Figure~\ref{fig:RecTD} there is no significant difference between the units. 
			Both units state with more than 75.0\% agreement that they usually recognize it when TD items are contracted on.

			An obvious difference can be seen in Figure~\ref{fig:CompareQvO}. 
			The units are asked whether they compare different solution alternatives and whether they estimate respective implementation costs, principal, and interest rates for taking on TD. 
			%If these comparisons are made, a conscious decision to take on TD can be assumed (RQ2.2).
			To answer RQ2.2, we assume conscious decisions to take on TD, if these comparisons are made.
			75.0\% of the participants of the observed unit agree to compare quick and optimum (or more) solutions and even calculate implementation cost for this comparison. 
			For the comparison unit 80.0\% survey participants do not agree to this which means they usually do not compare different solutions and estimate implementation costs. 
			Many participants of the observed unit agree that they also estimate principal (46.7\%) and interest costs (33.3\%), while no participants of the comparison unit do this.
			
			These observations lead to the following research hypothesis (H\textsubscript{R}) in comparison to respective null hypothesis (H\textsubscript{0}):
			
    		\textbf{H\textsubscript{R.1}:} The observed unit compares quick and optimum solutions more often than the comparison unit. 
    		(H\textsubscript{0.1}: The observed unit does not show a different behavior than the comparison unit in terms of comparing quick and optimum solutions.)
    		
    		\textbf{H\textsubscript{R.2-4}:} The observed unit estimates the implementation (and principal and interest) costs for different solutions more often than the comparison unit. 
    		(H\textsubscript{0.2-4}: The observed unit does not show a different behavior than the comparison unit in terms of estimating the implementation (and principal and interest) costs for different solutions.)

% 			\begin{itemize}
%     			\item H\textsubscript{R.1}: The observed unit compares quick and optimum solutions more often than the comparison unit. 
%     			(H\textsubscript{0.1}: The observed unit does not show a different behavior than the comparison unit in terms of comparing quick and optimum solutions.)
%     			\item H\textsubscript{R.2-4}: The observed unit estimates the implementation (and principal and interest) costs for different solutions more often than the comparison unit. 
%     			(H\textsubscript{0.2-4}: The observed unit does not show a different behavior than the comparison unit in terms of estimating the implementation (and principal and interest) costs for different solutions.)
% 			\end{itemize}
			
% 			\begin{itemize}
% 			\item H\textsubscript{R}: Units adopting this framework usually compare quick and optimum (or more) solutions. (H\textsubscript{0}: Units adopting this framework usually do not compare quick and optimum (or more) solutions.)
% 			\item H\textsubscript{R}: Units adopting this framework estimate the implementation costs for different solutions. (H\textsubscript{0}:  Units adopting this framework do not estimate the implementation costs for different solutions.)
% 			\end{itemize}
			
			\begin{table} 
			    \centering
        		\caption{Significance of H\textsubscript{R}-hypothesis}
        		\label{tab:MannWhitney}
        		\begin{tabular}{llc}
        			\toprule
        			H\textsubscript{R} 	& variable & exact significance\\  
        			\midrule 
        			H\textsubscript{R.1} & comparison		    & 0.058 \\ 
        			H\textsubscript{R.2} & implementation cost  & 0.048 \\ 
        			H\textsubscript{R.3} & principal cost	    & 0.002 \\
         			H\textsubscript{R.4} & interest cost 		& 0.015 \\ 
         			\bottomrule
        		\end{tabular}
        	\end{table}
			
			Table \ref{tab:MannWhitney} shows the significances of these hypotheses that were measured with the Mann-Whitney U-Test. 
			The exact significance for H\textsubscript{R.2-4} hypotheses is below 0.05 therefore these hypothesis can be accepted as they are significant to the 5\% significance level. 
			This means the possibility that these hypotheses are wrong is lower than 5\%.
			The exact significance for H\textsubscript{R.1} hypothesis is 0.058 which means the possibility that this hypothesis is wrong is only a little higher than 5\%.
			It can be accepted to the 10\% significance level.
			In summary, this means that the adoption of the framework leads to the named effects of a more conscious comparison of different solutions options and their respective costs.

			To answer RQ3.1 and 3.2, the participants are asked what benefits they expect (comparison unit) or assessed (observed unit) from the comparison of different solution option and therefore of the conscious contraction of TD.
			Figure~\ref{fig:benefits} shows that both units mostly agreed that TD could be prevented by this comparison. 
			Especially the observed unit assessed that discussions were led more rational (82.4\% agreement) and more rational decisions were made (76.5\% agreement). 
			Timely repayment (RQ2.3) of TD was not expected by the comparison unit but assessed by many participants of the observed unit (64.7\% agreement). 
			This may partly be due to a misunderstanding that the observed unit was unconsciously assessing the benefits of the entire framework and not just the benefits of the comparison of solution options.
			There is no obvious difference between the units regarding these questions which means no hypotheses can be derived. 

			To answer RQ3.3, the participants are asked whether the effort for comparing and estimating the different solutions is justified by the gained or expected benefits.
			Both units mostly think that the effort is justified (\textgreater80.0\% agreement). 
			There is also no obvious difference between the units as can be seen in Figure~\ref{fig:EffortJustified}. No hypotheses are derived. 
		
 %---------------------------------------------------------------------------------
			\subsubsection  {Correlations regarding the benefits of the framework} 
			
			As shown in the previous section, the most significant difference that arise by adopting the framework is the ongoing comparison between different solutions for every new ticket. 
			As pointed out in Section \ref{section:SurveyEvaluation}, the benefits of the framework cannot be assessed by the comparison unit.
			To identify the benefits of the comparison, we therefore analyze correlations between all survey participants who state that they compare different solutions and the benefits they therefore assessed. 

		 	Table \ref{tab:Correlations} shows all relevant and significant correlations that can be found.

			\begin{table} 
			    \centering
				\caption{Correlations of effects vs. benefits}
				\label{tab:Correlations}
	
				\subtable[Comparing solutions vs. assessing TD prevention and rational decisions as benefit]
					{	\label{tab:corComp}
						\begin{tabular}{llll|r}
						\toprule
											&			& \multicolumn{2}{c|}{comparing solutions} 	& $\phi$-coeff\\
											&			& not applies	&applies					& signif.\\ 
						\midrule 
						TD prevention 		&not applies& 42.9\% 			& 7.7\% 				& 0.419\\ 
											&applies 	& 57.1\% 			&\textbf{92.3\%}  		& 0.061\\ 
						\midrule 
						rat. decisions 	&not applies& 66.7\%			& 7.7\%  				& 0.623\\ 
									       	&applies 	& 33.3\% 			&\textbf{92.3\%}  		& 0.007\\ 
		
			 			\bottomrule
						\end{tabular}

					}

				\subtable[Estimating the implementation costs of different solutions vs. assessing rational decisions and rational discussions as benefit]
					{	\label{tab:corEstImpl}
						\begin{tabular}{llll|r}
						\toprule
											&			 & \multicolumn{2}{c|}{estim. impl. costs}	&  $\phi$-coeff\\
											&			 & not applies	& applies				& signif.\\ 
						\midrule 
						rat. discussions&not  applies& 50.0\% 		& 8.3\% 				& 0.471\\ 
										&applies  	 & 50.0\% 		&\textbf{91.7\%}  		& 0.035\\ 
						\midrule 
						rat. decisions 	&not  applies& 57.1\% 		& 8.3\%  				& 0.535\\ 
										&applies 	 & 42.9\% 		&\textbf{91.7\%} 		& 0.020\\ 
			 			\bottomrule
						\end{tabular}
					}

				\subtable[Estimating the principal costs vs. assessing rational decisions and rational discussions as benefit]
					{	\label{tab:corEstPrin}
						\begin{tabular}{llll|r}
						\toprule
											&			 & \multicolumn{2}{c|}{estim. princ. costs}	&  $\phi$-coeff\\
											&			 & not applies	& applies				& signif.\\ 
						\midrule 
						rat. discussions&not  applies& 41.7\% 		& 0.0\% 				& 0.456\\ 
										&applies  	 & 58.3\% 		&\textbf{100.0\%}  		& 0.049\\ 
						\midrule 
						rat. decisions 	&not  applies& 45.5\% 		& 0.9\%  				& 0.495\\ 
										&applies 	 & 54.5\% 		&\textbf{100.0\%} 		& 0.037\\ 
			 			\bottomrule
						\end{tabular}
					}
				% \subtable[Comparing solutions vs. recognizing TDs while contracting it]
				% 	{	\label{tab:corCompRec}
				% 		\begin{tabular}{lllll}
				% 			\toprule
				% 							&				& \multicolumn{2}{c}{comparing solutions} &\\
				% 							&				& not applies	&applies		& signif.\\ 
				% 			\midrule 
				% 			recognizing TDs & not applies 	& 50.0\% 			& 0.0\%  			&\\ 
				% 							& applies 		& 50.0\% 			&\textbf{100.0\%} & 0.007\\ 
				%  			\bottomrule
				% 		\end{tabular}
				% 	}	
			\end{table}

			It is important to point out that a correlation between two variables usually lead to two possible interpretations. One interpretation could be that variable A is the cause and B is the consequence and vice versa. In the following part only the most likely and reasonable interpretations are presented.

%Platz sparen durch Entfernung von Itemize		
		%	\begin{itemize}
		%		\item 
				\textbf{Table~\ref{tab:corComp}} shows the correlation between participants that compare sub-optimal and optimal (or more) solutions and participants that have assessed the benefits of TD prevention and rational decision making. 
				The correlation for TD prevention is medium strong (0.419) but with an asymptotic significance of 0.061 only significant to the 10\% significance level. 
				The correlation for rational decision making is strong (0.623) and with an asymptotic significance of 0.007 significant even to the 1\% significance level. 
				This can be interpreted to mean that comparing different solution options leads to the named benefits. 
				The possibility that this correlation does not exist, and our assumptions are wrong is higher for TD prevention ( \textless10\%) than for rational decision making (\textless1\%).
				
		%		\item 
				\textbf{Table~\ref{tab:corEstImpl}} shows the correlation between participants that are estimating the implementation costs of different solutions and participants that have assessed the benefits of rational discussions and rational decision making 
				These correlations are both medium strong (0.471 and 0.535) and significant to the 5\% significance level (0.035 and 0.02).
				This can be interpreted to mean that estimating the implementation costs of different solutions leads to the named benefits. 
	
		%		\item 
				\textbf{Table~\ref{tab:corEstPrin}} shows the correlation between participants that are estimating the principal costs and participants that have assessed the benefits of rational discussions and rational decision making 
				These correlations are both medium strong (0.456 and 0.495) and significant to the 5\% significance level (0.049 and 0.037).
				This can be interpreted to mean that estimating the implementation costs of different solutions leads to the named benefits.
		%	\end{itemize}
			
%			Other correlations, e.g. between comparing solutions and rational discussions, were not significant to the 5\% level and for this reason are not presented here.
 %%---------------------------------------------------------------------------------
			\subsubsection  {Open Questions}  
			Most comments were just made one time and are therefore just single opinions. 
			There were two points that were mentioned three times each. 
			
			First, ``TD is a topic'' expresses that it is already helpful to talk about the topic of TD and not to ignore it which is related to the goal to raise the of the overall TD awareness.
			This was pointed out by one participant: 
			``\textit{I actually see the main advantage of a conscious handling of the topic in the matter as such, because otherwise it is a topic that is almost ignored in everyday development}''. 
			
			The second code is ``Sometimes TD is OK'' where the participants were considering situations in which cases it is reasonable to keep TD, e.g. in legacy systems. 
			The following was stated by another participant: 
			``\textit{If a system is to be replaced, then I consider technical debts to be justifiable if they disappear after 1-2 years anyway}''. 
			
			One other point was mentioned two times which is ``Evaluation is difficult''. 
			In research this topic is known as TD prioritization problem. 
			It was phrased by one participant as follows: 
			``\textit{Even if the ``textbook'' says that the current implementation does not correspond to the architectural specifications, it still seems to be a personal opinion that decides on the fundamental correctness of the specification}''. 

			Two relevant drawbacks were mentioned by one participant, respectively. 
			One participant stated that the distinction of the categories and the separation of their responsibilities may lead to a separation of the team.
			``\textit{Although we keep saying that we are a team, we often stay separate and the different tickets in particular help make the differences between the team parts visible}''.
			Another participant thought that the framework ``\textit{(slightly) contradicts the agile principle of providing a functioning solution with company value as quickly as possible. 
			At the same time, however, it levels the stable ground so that we can continue to work on new requirements quickly and without production errors. The advantage outweighs}''.

	\section{DISCUSSION}
	\label{section:discussion}

    \subsection  {Framework application}
		\label{subsection:disRq}
		
	%	The basis of this work and the respective observations is the framework for managing TD as described in Section~\ref{section:framework}. 
	%	The RQ's are dedicated to feasibility, effects and benefits of this framework.

		\textit{\textbf{RQ1.1/RQ1.2:} Do practitioners find the framework reasonable?
		Are the processes of the framework feasible in practice?}
		
		Figures~\ref{fig:RecordingRes} to ~\ref{fig:ProcessingProcWork} show that there is very little disagreement. 
		Most survey participants find the framework reasonable and think that it does work as indented. 
		Only few participants think that it must at least be optimized.
		The ticket statistics (see Section \ref{section:TicketStatistics}) also indicate that the tickets are recorded and processed as intended.
		Furthermore, the framework is already used in the IT unit for more than two years.
		It can therefore be concluded that in an overall view the framework is feasible in practice and that the following findings based on this framework are valid.
			
    \subsection  {Framework effectiveness}
        \textit{\textbf{RQ2.1:} Does the framework lead to raised awareness for the contraction of TD?}
        
        Figure~\ref{fig:RecTD} shows that most participant are aware of it when they contract TD.
        Yet, there seems to be no difference between the observed and the comparison unit.
        The answers to the open questions support the finding that the awareness was raised considerably (``TD is a topic'') and that the participants started to consider when to take on or keep TD and when not to (``Sometimes TD are OK'').
        
        \textit{\textbf{RQ2.2:} Are the TD items taken on more consciously when using the framework?}
        
        Figure~\ref{fig:CompareQvO} reveals that in comparison the observed unit shows a significantly different behavior than the comparison unit. 
        The members of the unit using the framework are comparing different solutions, i.e. optimal and sub-optimal, and their costs before deciding for one of them.
        They may still take on TD by choosing the sub-optimal solution, but this is a conscious decision.
        
        \textit{\textbf{RQ2.3:} Are the TD items paid back timely?}
        
        If TD has to be contracted it is mostly paid back timely which can be seen in Figure~\ref{fig:TDTbyTime}.
        Even though not all TD tickets may be paid back during their respective projects as intended by the framework. 
        Figure~\ref{fig:benefits} shows that two thirds of the participants of the observed IT unit perceive that TD items are paid back timelier. 
		Nevertheless, not all TD tickets are paid back and a third of the participants are still dissatisfied with the timely repayment.
		
    \subsection{ Framework benefits}
        \textit{\textbf{RQ3.1/RQ3.2:} Can TD be prevented by the adoption of the framework?
        Are there other benefits arising from the adoption of the framework?}
        
        The most obvious effect of the framework is the conscious contraction of TD by comparing different solution options.
        In Figure~\ref{fig:benefits} the benefits that arise by these are evaluated. 
        Both units assessed or expected benefits of this comparison. 
        
        Furthermore, the correlations between the benefits and participants that did this comparison are significant. 
        This paper was able to show that the comparison led to more rational discussions and decisions. 
        The error probability for this assumption is lower than 5\%. 
        The prevention of TD was assessed with 6.1\% only to the lower 10\% significance level which means that it is only a little less likely that this observation is valid.
        These benefits can therefore be referred to the use of the framework.
        
        \textit{\textbf{RQ3.3:} Do these benefits justify the additional effort?}
        
		Finally, Figure~\ref{fig:EffortJustified} shows that most participants think that the additional effort for comparison and cost estimation is justified and thus backs up the usefulness of this framework.
		
		\subsection  {Threats to Validity}
		\label{subsection:disTtV}

		\textbf{Construct Validity}:
		To enhance construct validity, we questioned and compared two units one of which did adopt the framework and one that did not use any method for managing TD. 
		To ensure comprehensibility and to optimize the questionnaire, we did pilot tests. 
		It could be seen as a problem that the survey did not directly but only in open questions ask for drawbacks. %of the framework. 
		The questions whether the effort for the benefits is justified should have remedied parts this problem. 
		Additionally, we not only evaluated the questionnaire but also the ticket statistics which further improves the construct validity, e.g. when evaluating the timely repayment of TD. 
	
		\textbf{Conclusion Validity}:
		The paper was able to show statistically significant differences between the observed and the comparison unit. 
		We used SPSS and standard statistic techniques for the evaluation of the significance of the findings. 
		On the other hand, the sample was small.
		The comparison unit was considerably smaller than the observed unit and had a slightly different organization. 
		By choosing a unit that was led by the same unit manager this threat was minimized. 
		The survey response was only about 50\% of the respective units which can lead to deviations in the evaluation. 
		
		\textbf{Internal Validity}:
		Some valuable effects were observed, but it is possible that the same effects could have been observed in other IT units. 
		This threat was minimized by the comparison with a unit that did not use the framework. 
		Nevertheless, just the management of TD could have led to a better developer morale like presented in \cite{Besker2020a} and therefore to good ratings. % for the framework.
		This threat can and should be minimized by replicating the study in other constellations. 
		As correlations do not directly lead to causalities, the mentioned interpretations of the correlations should be validated, e.g. by follow up interviews.

		\textbf{External Validity}:
		It is an important benefit of this paper that the framework was developed and tested in an industry environment and that the framework has been used for a long time. 
		Nevertheless, this was just a case study and can be biased. 
		The framework adoption and the study should be replicated in other units and companies, especially in other economic sectors and countries. 
		%This study can be replicated using the provided information in this paper.
		%The data for replication and verification can be downloaded online\footnote{\url{https://doi.org/10.5281/zenodo.4616485}}. 
		By providing the framework information in this paper and the questionnaire and raw data\footnote{\url{https://doi.org/10.5281/zenodo.4616485}} the study can be replicated to back up the findings.
	
		\subsection  {Related Work}
		\label{subsection:disRelW}
% 		Ward Cunningham first described TD: ``\textit{Shipping first time code is like going into debt. A little debt speeds development so long as it is paid back promptly with a rewrite}'' (\cite{Cunningham1992}). 
% 		TD tickets as described in this paper adhere to this original description as a decision for a sub-optimal solution to speed up development. 
% 		The framework further provides a mechanism to pay back the debt. 
		
% 		In contrast, with maintenance tickets another type of tickets is introduced for maintenance issues that are not contracted intentionally. This categorization follows the idea of intentional and unintentional or deliberate and inadvertent TD as presented in \cite{Fowler2009} and \cite{McConnell2008a}. Frank Buschmann \cite{Buschmann2011} finds that ``\textit{the metaphor works best when technical debt is intentional}''.
	%	Avgeriou et al. \cite{Avgeriou2016a} describe this distinction as follows: ``While the conceptual roots of technical debt imply an idealized, deliberate decision-making process and rework strategy as needed, we now understand that technical debt is often incurred unintentionally and catches software developers by surprise''

\textbf{TD Prevention}:
	    In a systematic mapping study in 2015 Li et al. \cite{Li2015} identified eight TD activities one of them being TD prevention which was with seven studies between 1992 and 2013 also one of the obviously understudied activities. 
		In 2016 Yli-Huumo et al. \cite{Yli-Huumo2016} find TD prevention to be ``\textit{one of the most influential activities of the eight TDM activities that a development team can conduct}''. 
		In 2018 Rios et al. \cite{Rios2018} started to fill this gap by indicating that TD can be prevented, and that TD prevention is a worthwhile research topic. 
		In their follow-up work \cite{Freire2020}, they identified actions and impediments to TD prevention by evaluating a questionnaire filled by practitioners. 
		The work did not provide ideas how to implement these actions in practice or how to deal with the impediments. 
		%The participants were asked for an TD example and how the TD could have been prevented. 
		%Unfortunately, they were not asked why they did not prevent it, which may have led to the root-cause for missing TD prevention like e.g. time pressure. 
		
\textbf{Timeline Problem}:
		As already mentioned in Section~\ref{section:introduction} one of the main root causes for TD to be found in all papers that deal with TD causes is the problem of tight timelines (e.g. \cite{Freire2020,Rios2018,Martini2014,Verdecchia2020,Avgeriou2016a}). %,Buschmann2011,Malakuti2020
		To the best of the authors knowledge no papers contribute a solution to this problem.
		By using TD tickets, this framework fills this gap and provides a practical method that has impact on two of the actions listed in \cite{Freire2020}: (I) ``\textit{Following a well-defined project planning}'' by integrating TD management into project management, (II) ``\textit{Having an effective team}'' with ``\textit{good communication on the team}'' as shown by the impact the framework has on team discussions. 
		Codabux et al. \cite{Codabux2014} also list different strategies to prevent TD, e.g. ``\textit{Education and Training}'', `\textit{`Pair Programming}'', ``\textit{Conformance to process and standards}'', but they also do not provide a solution on how to stick to the strategies under time pressure.
		This framework puts the responsibility for TD ticket repayment to the business analysts and therefore project management. 
		Thereby, this paper proposes a solution to the problem of ``\textit{Split of budget in Project budget and Maintenance budget boosts the accumulation of debt}'' (\cite{Martini2014}). 

		%Ist schon unten im letzten Abschnitt erwähnt - sollte reichen ...
		%Guo et al. \cite{Guo2016a} also propose a framework for managing TD which focuses on TD identification, TD measurement and TD monitoring while this framework focuses TD repayment and TD prevention. Getting a better overview what is within the scope of TD monitoring is a secondary goal of this framework. Furthermore Guo et al. provide an idea for TD prioritization. It could be interesting future work to evaluate whether the frameworks can be integrated to a more holistic approach.

% ## ??? ##	
% 		In terms of a management strategy as introduced by Verdecchia et al. \cite{Verdecchia2020}, this framework represents an active management strategy. It might serve as a practical example to the subcategory ``\textit{technical credit}'' which was missing in the original study, where this category was deemed a mere theoretical one. 
% 		This framework therefore really presents a new approach, that could not be found so far.
		
\textbf{TD Awareness}:
		One important effect of this framework is the raised awareness for contracting TD.
		Kruchten \cite{Kruchten2012a} already mentioned the importance of awareness: ``\textit{The first step is awareness: identifying debt and its causes}''. 
		In a survey with more than 1,800 participants, Ernst et al. \cite{Ernst2015} find that that 79\% of the participants strongly agree that ``\textit{Lack of awareness of TD is a problem}''. 
		%Verdecchia et al. \cite{Verdecchia2020} also mention that ``\textit{people’s personal drive, skill set, and awareness can influence ATD}'', where ATD is the abbreviation for architectural TD.
		Tonin et al. \cite{Tonin2017} researched the effects of TD awareness by conducting a classroom study. 
% 		The students were informed about the concept of TD and should use TD boards during a project. 
% 		The effects of this work were a changed attitude, more discussions, and a more conscious contraction of TD. 
% 		These effects are similar to the findings in this paper. 
		In contrast to that this paper presents a method to raise the awareness %other than just talking or learning about it. 
		%Also, it is able to raise the awareness 
		not only for developers but also for business analysts and managers. 
		Finally, this paper presents an evaluation in industry, which provides a higher external validity.

\textbf{TD Management Case Studies}:
% 		Other case studies concerning TD management are \cite{Malakuti2020}, \cite{Martini2015} and \cite{Guo2016a}. 
% 		They present research findings in the same area as this study as they try to tackle impact of TD in industry.
% 		Yet, in \cite{Martini2015} Martini et al. follow a more organizational solution to TD management and in \cite{Malakuti2020} Malakuti and Ostroumov describe a specific project but are still missing a working solution. 
		Other case studies concerning TD management are for example \cite{Martini2015} and \cite{Guo2016a}. 
		They present research findings in the same area as this study as they try to tackle impact of TD in industry.
		Yet, in \cite{Martini2015} Martini et al. follow a more organizational solution to TD management.
		Guo et al. \cite{Guo2016a} focus their framework on the TD identification, TD measurement and TD monitoring. Furthermore, Guo et al. provide an idea for TD prioritization. It could be interesting future work to evaluate whether the frameworks can be integrated to a more holistic approach.
		%Furthermore, \cite{Malakuti2020} and \cite{Martini2015} only refer to large-scale organizations while this study supports an approach that is feasible independent of the organization size.
	
	\section{CONCLUSION AND FUTURE WORK}
	\label{section:conclusion}

         %\subsection  {Conclusion}
    	With this paper we present and evaluate a framework that includes TD management in project management by so-called TD tickets. 
        TD tickets hand over the responsibility for TD contracted during a project to the project management of the same project. This creates a feedback loop of team and management behavior that leads to TD prevention.
    	As it was developed and extensively used in practice, the framework shows a high external validity and provides the research community with and insight into the practical handling of TD.
    	
    	The adoption of the framework led to a higher TD awareness and conscious contraction of TD in the case company. 
    	This on the other hand leads to the prevention of TD, more rational discussions between the unit members and more rational decisions. 
    	Furthermore, the TD items taken on during a project are part of the project and will be paid back timelier. 
    	While the idea of maintenance tickets and projects is not new, the idea of TD ticket is a novel approach that could not be found in research literature to the best of the author's knowledge.
    	
    	For industry, this paper presents a framework that can be adopted, refined, and adapted to their own needs. 
    	The idea of including TD management in project management can be an incentive to develop a similar approach. 
    	Also, the importance of raising TD awareness and contracting TD intentionally by taking the time to compare different solution options was presented in this paper. 
    	The advantages shown by this paper can be helpful when negotiating options to handle TD with the IT management.
    
    	For researchers, this paper provides insights to procedures for managing TD developed in industry. 
    	The paper indicates the importance %of feasible and structured solutions for handling and preventing TD. 
    	%This means 
    	to provide a solution for handling and preventing TD even under the pressure of a tight timeline. 
    	The paper also confirms the problem that the responsibility for TD needs to be transferred to the whole unit and cannot be solved by developer teams alone. 
    	The framework presented in this paper provides a solution for these problems. 
    	
    %	For future work, it can be worthwhile to evaluate if the comparison of solutions leads to better compromises that may not be the optimal solutions but one that may not need to be refactored either.
    %	Furthermore, the importance of raising the awareness for TD in the whole unit and not only the development team is an understudied area, and more ideas should be developed and evaluated.
    	For future work, the framework could be complemented by other TD activities, e.g. TD prioritization, TD monitoring and TD identification, and their impact could be evaluated. Especially ideas to TD Prioritization may help to optimize the repayment phase as this was mentioned as a problem.
    	The management of deconstruction tickets in particular is an interesting and promising research topic and should be evaluated in future. 
    	Timely repayment of TD is sometimes not performed due to low priority or missing awareness of competing goals. Further research with respect to project management and business value estimation is necessary. 
   % 	The timely repayment got negative feedback from some of the participants of the observed IT unit.
    %	It should be evaluated in what way this discontent could be managed, e.g. by prioritization or deferral.
	    Finally, the prioritization of maintenance tasks including TD should be assessed between different application domains and industries.
	    
    	%One idea is to only repay high priority tickets during the project. %It is then another task to find a way to manage the deferred tickets. %, e.g. by linking the repayment to the next change of the affected part of the system.
    	%This will be part of our future work.

%%
%% The acknowledgments section is defined using the "acks" environment
%% (and NOT an unnumbered section). This ensures the proper
%% identification of the section in the article metadata, and the
%% consistent spelling of the heading.

\section*{Acknowledgment}
We would like to thank {\it Gruner+Jahr GmbH - Media Sales Services} and {\it - Smart Business Solutions} for their contributions to the framework, for completing the questionnaire and for their support of this paper.

%%
%% The next two lines define the bibliography style to be used, and
%% the bibliography file.
%\section*{References}
% \bibliographystyle{ACM-Reference-Format}
% \bibliography{TechDebt}

\bibliographystyle{IEEEtran}
\bibliography{IEEEabrv,TDAwareProjectManagement}

\end{document}